\documentclass[twocolumn]{aastex631}
\usepackage{booktabs}
\usepackage{array}
\usepackage[caption=false]{subfig}
\usepackage{comment}

\newcommand\s{{\rm~s}}

\newcommand\hst{{\it HST}}
\newcommand\chandra{{\it Chandra}}
\newcommand\sax{{\it BeppoSAX}}
\newcommand\suzaku{{\it Suzaku}}
\newcommand\asca{{\it ASCA}}
\newcommand\rxte{{\it RXTE}}
\newcommand\nustar{{\it NuSTAR}}
\newcommand\einstein{{\it Einstein}}
\newcommand\exosat{{\it EXOSAT}}
\newcommand\ginga{{\it Ginga}}
\newcommand\astrosat{\textit{AstroSat}}
\newcommand\gaia{\textit{Gaia}}
\newcommand\xmm{\textit{XMM-Newton}}

\shorttitle{UVIT deep field around IC~4329A}
\shortauthors{Ganguly et. al.}
\graphicspath{{./}{Figures/}}

\begin{document}
\title{\astrosat{}/UVIT far and near UV deep field around IC~4329A}

\author{Piyali Ganguly}
\affiliation{Inter-University Centre for Astronomy and Astrophysics (IUCAA), PB No.4, Ganeshkhind, Pune-411007, India\\}

\author{Priyanka Rani}
\affiliation{Inter-University Centre for Astronomy and Astrophysics (IUCAA), PB No.4, Ganeshkhind, Pune-411007, India\\}

\author{Gulab C. Dewangan}
\affiliation{Inter-University Centre for Astronomy and Astrophysics (IUCAA), PB No.4, Ganeshkhind, Pune-411007, India\\}



\begin{abstract}
We present high-resolution near-ultraviolet (NUV) and far-ultraviolet (FUV) deep imaging  of the field around the Seyfert galaxy IC~4329A based on five observations performed with the Ultra-Violet Imaging Telescope (UVIT), onboard \astrosat{}.  The long exposures of 82.9 ks in NUV (N245M; $\lambda_{mean}=2447$\AA;  $\Delta\lambda = 270$\AA) and 92.2~ks in FUV (F154W;  $\lambda_{mean} = 1541$\AA; $\Delta\lambda=380$\AA) bands constitute the deepest observations with $5\sigma$ detection limits of AB magnitudes $m_{NUV}= 26.2$ and $m_{FUV} = 25.7$. Leveraging UVIT's excellent angular resolution (FWHM $\sim 1.2-1.8\arcsec$), we performed a detailed analysis of the IC~4329A field and detected (above 5$\sigma$ significance level) a total of 4437 and 456 sources in the NUV and FUV bands, respectively. A large number of these detected sources were unknown previously.
We performed astrometry and photometry on all detected sources. By cross-matching our catalogue with \gaia{}-DR3 and \xmm{} DR12 catalogues, we found 651 optical and 97 X-ray counterparts of our sources. Additionally, we explored UV variability of point sources, identifying 28 NUV sources as variable with a significance above the $2.5\sigma$ level. Of these, only three sources exhibited variability in the FUV band. Utilising the NUV and \textit{Gaia} fluxes, we determined that two previously catalogued white dwarf candidates are misclassified. Furthermore, we highlight galaxies with atypical morphology, including ring-like structures, multiple compact central sources,  bifurcating spiral arms, etc. Follow-up optical spectroscopy and multi-wavelength observations are imperative to further investigate the nature of the sources within this field.
\end{abstract}



\section{Introduction} \label{sec:intro}

Deep astronomical imaging observations provide a wealth of information on a huge  number of sources with a wide range in redshifts, morphological types, flux and magnitudes, colors, etc.  These data help us to identify and characterize different classes of astronomical sources, luminosity functions and  evolution of different classes of sources over cosmic time. 
Deep imaging  enables us to study the most distant galaxies, which are too faint to be seen via traditional imaging techniques. By imaging these faint galaxies, we can better understand the early universe and its evolution \citep[see e.g.,][]{2008ApJ...678..751R,2009ApJ...701.1765M,2010ApJ...709L.133B}. 
Multi-band photometric surveys help us to construct spectral energy distributions (SEDs) and investigate a variety of emission processes responsible for a diverse set of astrophysical phenomena occurring in different types of sources. SED studies of galaxy samples have been used to estimate photometric redshifts, stellar mass and star formation rates.

Optical astronomy has a long history of deep field observations. Two of the most significant deep imaging surveys ever conducted are the Hubble Ultra Deep Field (HUDF) \citep{2006AJ....132.1729B}, and the Cosmic Assembly Near-infrared Deep Extragalactic Legacy Survey (CANDELS; \citep{2011ApJS..197...35G,2011ApJS..197...36K}).  The Hubble deep fields north and south (HDF-N and HDF-S) are the deepest and data-richest part of the celestial sphere that have provided the most comprehensive picture of the Universe. These deep fields discovered a large number of galaxies at high redshifts (up to $z=6$), and revealed that galaxy collisions and mergers were much more common at high redshifts than in the local Universe.  Observation of a large number of galaxies  at different stages of evolution also reveal galaxy evolution and cosmic history of star formation. Subsequent observations of HUDFs at IR,  X-rays and radio wavelengths have led to a wealth of information \citep[see e.g.,][]{2022ApJ...941..191L, 2020ApJ...901..168A, 2019ApJ...882..140B, 2023ApJS..268...64W}

\astrosat{}'s Ultra-Violet Imaging telescope (UVIT) has performed deep field observations that have yielded interesting results. \citet{Mondal_2023} performed deep UVIT observations of the  Great Observatories Origins Survey Northern (GOODS-N) field  (\astrosat{} UV Deep Field north—AUDFn) using FUV and NUV filters down to $3\sigma$ magnitude limits of $m_{AB} \sim 27.35$, $27.28$, and $27.02$ mag for a point source in the UVIT filters F154W, N242W, and N245M, respectively. They detected 16001 sources in the FUV band and 16761 in the NUV band, thus demonstrating high detection sensitivity of the UVIT. 
\citet{2022PASA...39...48M} performed deepest UVIT/FUV observations of the central field in the Coma cluster, and detected 1308 sources including galaxies with unusual FUV morphology.

We utilise the deep UVIT field acquired with the long \astrosat{} observations of a well known Seyfert 1 galaxy IC~4329A \citep{2021MNRAS.504.4015D}.  IC~4329A is the second-brightest type 1 AGN in
the Swift/BAT catalog \citep{2013ApJS..207...19B}, and is located at a redshift of $z = 0.016054$ \citep{1991AJ....101...57W}. This AGN lies at $14.8 {\rm~arcmin}$ away from the center of the Abell cluster A~3574.  This cluster is the most distant and  easternmost member of the chain-like Hydra-Centaurus supercluster \citep{1979ApJ...230..648C}. The brightest galaxy in the cluster is the giant lenticular galaxy IC~4329. This galaxy  has a similar redshift as that of IC~4329A, and lies at a projected distance of 59~kpc to the west. The two galaxies IC~4329 and IC~4329A appear to be part of a loose group of seven galaxies, and likely going through an interaction. Apart from IC~4329A, none of the other six group members show any indication of significant AGN activity. 
Due to the presence of the powerful AGN, the field has been observed with all major observatories including  \hst{}~\citep{1998ApJS..117...25M}, \einstein{}~\citep{1984ApJ...280..499P, 1989ESASP.296.1105H}, \exosat{}~\citep{1991ApJ...377..417S}, \ginga{}~\citep{1990ApJ...360L..35P},  \asca{}/\rxte{}~\citep{2000ApJ...536..213D}, \sax{}~\citep{2002A&A...389..802P}, \chandra{}~\citep{2004ApJ...608..157M}, \xmm{}~\citep{2007MNRAS.382..194N}, \suzaku{}~\citep{2016MNRAS.458.4198M}, and  \nustar{}~\citep{2014ApJ...788...61B}. 
 However, these observations have been utilised for detailed study of the AGN or the galaxy group.  Most of the above fields are also smaller in size than the $28{\rm~arcmin}$ diameter circular field acquired with the UVIT. Till now, studies on IC4329A have mostly been limited to pointed observations (such as, from \chandra{} and \xmm{}, or from \hst{}) and therefore, the field around it remained less explored. 
For example, the same IC4329A field has already been observed by Galaxy Evolution Explorer telescope \cite[GALEX;][]{2005ApJ...619L...1M} in FUV and NUV bands, but its inferior angular resolution (FWHM $\approx{5\arcsec}$ ) makes the data less effective for a deep field study. 
Thus, a lack of deep, high-resolution far-UV (FUV) and near-UV (NUV) imaging data remained the bottleneck in doing such a study.

\begin{figure*}
\gridline{
        \fig{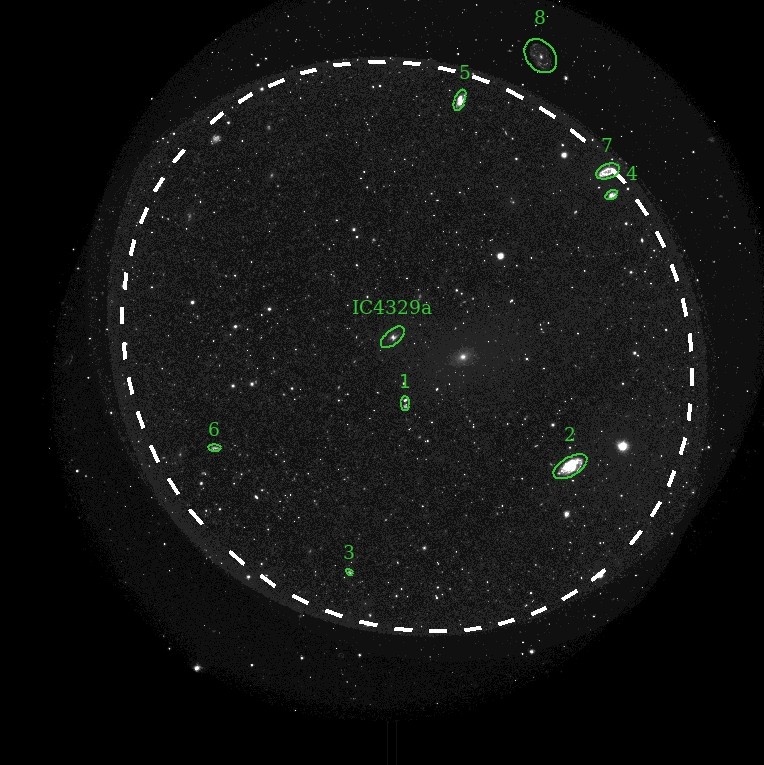}{0.5\textwidth}{}
        \fig{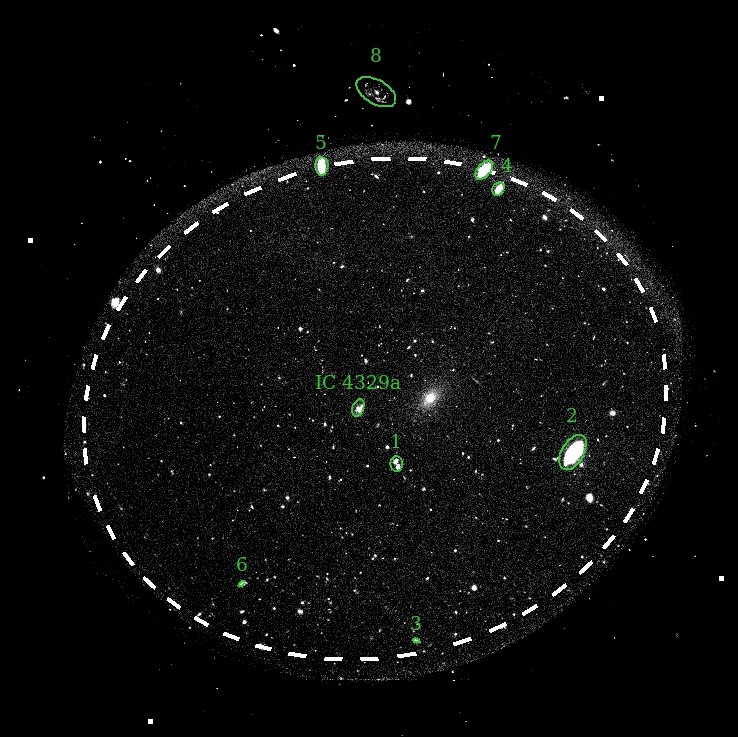}{0.5\textwidth}{}
        }
\caption{The merged UVIT images obtained from five observations of IC~4329a in the NUV N245M (left panel) and FUV F154W (right panel) bands. The part of the field enclosed by white colored dashed ellipse in each of the images is common to all five observations and only this part of the field has been used for making the catalogue. The common field of view in the NUV filter is centered at $\alpha$ = 207.31778 degrees and $\delta$ = -30.30287 degrees and measures 470 ${\rm{arcmin}}^2$ area and the common field of view in the FUV filter is  centered at $\alpha$ = 207.31706 degrees and $\delta$ = -30.30465 degrees and measures 423 ${\rm{arcmin}^2}$ area. The Seyfert galaxy IC~4329a and some other extended sources that are discussed in Sections~\ref{sect:int_sources} (sources 1--6 in the image) and \ref{sect:int_sources_2} (sources 7 and 8 in the image) are marked with green-coloured regions.
\label{fig:field}}
\end{figure*}

In this context, we use high-resolution  UV imaging observations (in FUV and NUV) of the IC4329A field, taken by Ultra-Violet Imaging Telescope \cite[UVIT;][]{2012SPIE.8443E..1NK,2017AJ....154..128T}. UVIT has an FWHM (full width at half maximum) resolution of  $\approx{1.2-1.8\arcsec}$, which is almost three times better than that of the GALEX. In this paper, we  present a catalogue of UV sources within this field and outline their positions, sizes, signal to noise ratios (SNR), magnitudes etc. Interestingly, some of these sources were not detected/resolved by GALEX and hence, have been identified for the first time, thanks to UVIT's superior angular resolution and deep observations. 
We organise the paper as follows. We describe UVIT observations, data reduction and generation of science images in Section~\ref{sect:data_reduction}. We characterize the FUV and NUV background levels and the point spread functions in Section~\ref{sect:analysis}. We also describe source detection, astrometry and photometry and present the source catalogue and cross-correlate the detected FUV and NUV sources with those known in the optical and X-ray bands in Section \ref{sect:analysis}. We have quoted all the source co-ordinates in epoch J2000 throughout this paper. We present our final catalogue, investigate the variability of the sources and also try to identify possible Active Galactic Nuclei candidate in our catalogue in Section \ref{sect:results}. We summarise main results of our paper in Section~\ref{sect:summary}.

\section{UVIT observations and data reduction}\label{sect:data_reduction}
The UVIT payload, on board \astrosat{}, houses two telescopes, each with a primary mirror of 375~mm in diameter and a field of view (FOV) of a circular region with  $28\arcmin$ diameter. The two telescopes constitute three channels: Far-UV (FUV; 130–180 nm), near-UV (NUV; 200–300 nm) and visible (VIS; 350–550 nm). 
The three channels utilise identical intensified CMOS imaging detectors, they differ in their photo-cathodes and window materials.
The detectors in FUV and NUV channels operate in photon counting mode and, are used for scientific observations, while the visible channel is primarily used to correct for drift of the satellite pointing and to achieve the required spatial resolution of the UV images.
 The FUV and NUV channels are equipped with a number of filters with different bandpasses for high resolution (FWHM$\sim 1.2-1.8\arcsec$) imaging of  $\sim 28{\rm~arcmin}$ diameter circular field. These channels are also equipped with slit-less gratings for low resolution spectroscopy.
 Lastly, the instrument is capable of operating all three channels in tandem. Details of the UVIT including calibration and performance can be found in \cite{2017AJ....154..128T,2020AJ....159..158T}. 

\begin{deluxetable*}{cccc}
\tablecaption{List of the UVIT observations.\label{Table:obslog1}}
\tablewidth{10pt}
\tablehead{
\colhead{Obs. Id. } & \colhead{Obs. Time} & \colhead{Filter name} & \colhead{Exposure Time}
\\
& & &(sec)
}
\startdata
9000001006 	& 04-02-2017    & F154W, N245M   & 18810.763, 17251.98\\  
9000001048 	&  24-02-2017  & F154W, N245M   &  19652.562, 19966.288\\ 
9000001118 	& 30-03-2017   & F154W, N245M   &  19748.1, 19863.619\\ 
9000001286 	&  12-06-2017  & F154W, N245M   &  18477.599, 19185.13\\ 
9000001340 	& 26-06-2017    & F154W, N245M   &  15497.699, 16941.51\\ 
\enddata
\end{deluxetable*}

\begin{deluxetable*}{cccccc}
\tablecaption{Details of the final merged UVIT images. Corresponding calibration data is taken from \citep{2017AJ....154..128T}
\label{Table:obslog2}}
\tablewidth{10pt}
\tablehead{
\colhead{Filter} & \colhead{${\lambda_{mean}}$} & \colhead{Zero point mag. } & \colhead{Unit conversion} & \colhead{Exposure time} & \colhead{5$\sigma$ detection limit } \\
& \colhead{(\AA)} & \colhead{(AB)}& \colhead{(erg sec$^{-1}$cm$^{-2}$\AA$^{-1}$)} & \colhead{(sec)} & \colhead{(AB magnitude)}
}

\startdata
F154W & 1541 & 17.778 & $3.55\times10^{-15}$ & 92186.7 & 25.75\\
N245M & 2447 & 18.50 & $7.25\times10^{-16}$ & 82885.9  & 26.16\\
\enddata
\end{deluxetable*}

\astrosat{}/UVIT observed the IC~4329A field five times during February to June 2017. We provide the details of these observations in Table~\ref{Table:obslog1}. The UVIT was operated in the Photon-Counting mode using
two broadband filters: FUV/F154W ($\lambda_{mean}$ = 1541\AA{}, $\Delta \lambda$ = 380\AA{}) and NUV/N245M  ($\lambda_{mean}$ =2447\AA{}, $\Delta \lambda$ = 280\AA{}). We obtained the {\tt level1} data  for the five observations 
from the \astrosat{} data archive\footnote{\url{https://astrobrowse.issdc.gov.in/astro_archive/archive/Home.jsp}}. We used the UVIT pipeline CCDLAB \citep{2017PASP..129k5002P, 2021JApA...42...30P} and processed the level1  FUV and NUV data from individual observations. This CCDLAB software performs corrections for image distortions, flat fielding, and spacecraft pointing drift. During data processing, we also checked for possible contamination from cosmic rays. A 4$\sigma$ thresholding scheme was applied to remove those frames that contain cosmic ray induced signal. Here, $\sigma$ = $\sqrt{I}$ and I is the median number of counts/frame calculated over all observed frames. We used the VIS channel images acquired every second, and derived drift series using bright stars. We applied the drift series to the FUV and NUV data and obtained drift-corrected orbit-wise centroid lists and images for each filter. We then aligned the orbit-wise drift-corrected data for each filter and merged them to obtain merged centroid list and image for each filter.

Using CCDLAB, we further optimised the
point spread function (PSF) in these deep images to correct for any higher-order drift residuals \citep{2021JApA...42...30P}, and generated centroid-lists and images free of cosmic-ray events. To further increase the signal-to-noise ratio (SNR) for a given filter, we merged images (after cross alignment) from all five observations. For merging all five observations we have used CCDLAB for NUV observations and for the FUV observations we have used the python package {\tt Astroalign}  \citep{BEROIZ2020100384}. Lastly, we performed astrometry of our field by using the Gaia EDR3 astrometric reference system \citep{2021A&A...649A...1G} through the built-in algorithm in CCDLAB software.
The astrometric position errors (rms) across the entire field of NUV and FUV images are $\sim$0.19$\arcsec$ and $\sim$0.60$\arcsec$,  respectively.

Since the fields covered by each observations are not exacly the same, we identified the common overlapping field as a large elliptical region for each filter. We show the merged images in the FUV and NUV filters in Figure \ref{fig:field} with the common elliptical fields marked by an white ellipse. This way, we ensured that the exposure time was uniform throughout the field of our interest and equal to the total observation time of all observations added. We also excluded outer rim regions of each field to avoid spurious detections. 
Thus, finally we are left with two science-ready images, one each for the FUV (Fig.~\ref{fig:field}) and the NUV (Fig.~\ref{fig:field}) filter. The net exposure time of the merged image in NUV is 82.9 ks wherein the same for FUV band is 92.2 ks. Each of these two final images is of the size of  4800$\times$4800 pixels (including padding of areas without coverage) at a plate scale of $\sim$0.416$\arcsec$. 
The details of the final FUV and NUV data are listed in Table~\ref{Table:obslog2}.

\section{Analysis}\label{sect:analysis}

\subsection{Background Characterization}
The determination of background levels in our final merged FUV and NUV images is critical in  detecting sources and measuring accurate flux, magnitude, signal-to-noise ratio (SNR) and other properties of the sources in the field. We measured the mean background level following a method similar to the one described in \citet{2020NatAs...4.1185S}.
To measure the average background counts, firstly we need to detect and mask out as many sources as possible, particularly the bright ones. For this purpose, we used the widely used photometry software {\tt `Source Extractor'} (or {\tt {SExtractor}}) tool \citep{1996A&AS..117..393B}.

We initially ran ${\tt {SExtractor}}$ to find all the sources that have at least 10 adjacent pixels with counts more than $2\sigma$ level above the background counts calculated by the ${\tt{SExtractor}}$. We achieved this by setting the input parameters DETECT\_MINAREA = 10 and DETECT\_THRESH = 2 with THRESHOLD\_TYPE as RELATIVE for both the fields.  This gave us two catalogue files listing bright sources in the FUV and NUV bands. We  used this initial source catalogue to mask out sources in order to measure the mean background level for each band. 

\begin{figure*}
\gridline{\fig{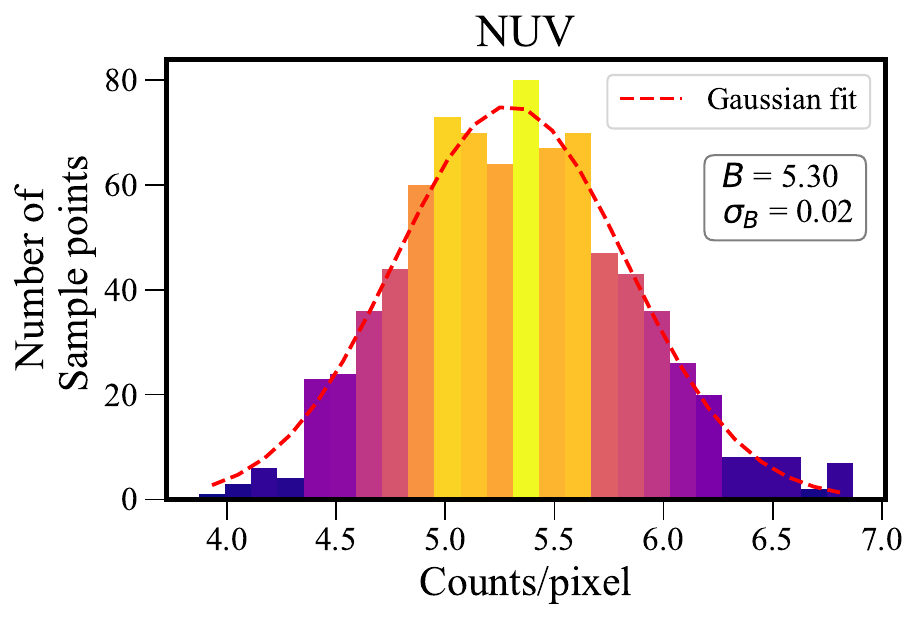}{0.5\textwidth}{a}
            \fig{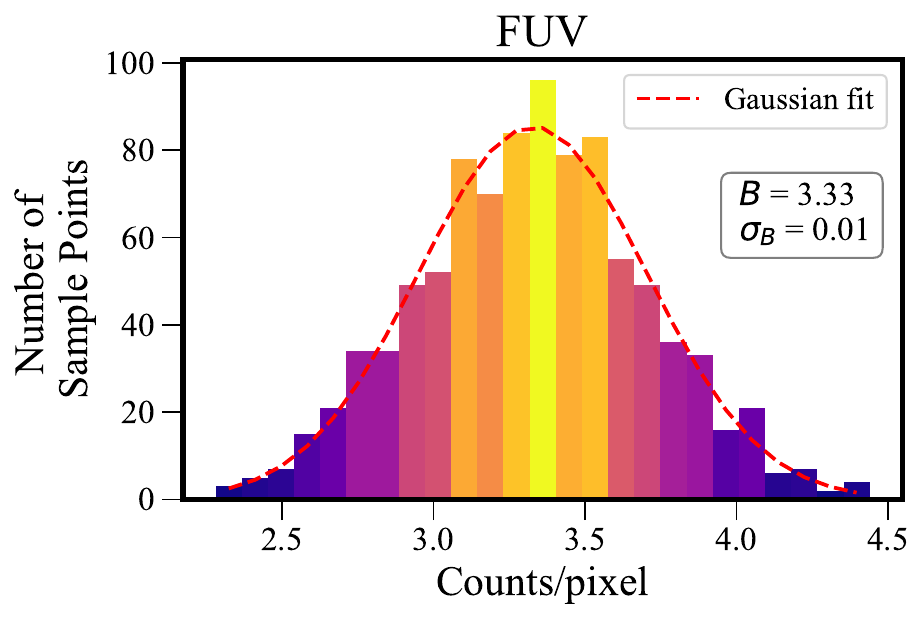}{0.5\textwidth}{b}
            }
\caption{The distributions of background counts per pixel for NUV (left panel) and FUV band (right panel) images. The distributions were generated using $\approx{1000}$ square boxes of size $5 \times 5 $ pixels placed across the image. The dotted curves show the best-fitting Gaussian profiles. We quote the mean background level ($B$) and its error ($\sigma_B$) in counts per pixel in the two panels for the NUV and FUV bands.
\label{fig:bkg_dist}}
\end{figure*}

First, we prepared a mask covering the area inside the Kron aperture for each detected source. For this, as well as all the photometric calculations presented in this paper, we have used Kron aperture calculated by $\tt{SExtractor}$ corresponding to KRON\_FACTOR = 2.5 which includes $94\%$ flux for a given source  \citep{1980ApJS...43..305K}. On this masked image we randomly put 1000  boxes each of $5\times5$ pixels throughout the whole image and measured the average counts per pixel for each box. In this calculation, we excluded any box that partially or entirely overlapped with the masked pixels or any box that was outside the common Field of View covered by five observations as shown in Figure~\ref{fig:field}.

We measured the total counts in each of these boxes and generated histograms of  counts per pixel. In order to reduce the possible contribution from any undetected faint source, we applied a sigma-clipping algorithm on this distribution iteratively. In each step, we calculated the mean and the standard deviation ($\sigma$), and excluded those boxes with counts above the $3\sigma$ level. We repeated this procedure  till the convergence i.e., no boxes left with counts more than $3\sigma$ level.  

We calculated the mean and the standard deviation of this sigma-clipped sample by fitting a Gaussian distribution as shown in Figure~\ref{fig:bkg_dist}. The  mean FUV and NUV background level and the standard deviation of the mean as measured from the Gaussian fit are $b = 6.39 \times 10^{-5} \pm 2.22 \times 10^{-7}{\rm~counts~s^{-1}~pixel^{-1}}$ for the NUV and $b = 3.61 \times 10^{-5} \pm 1.34 \times 10^{-7}{\rm~counts~s^{-1}~pixel^{-1}}$ for the FUV band.

\subsection{Point spread function (PSF)} 
\begin{figure}[ht!]
\plotone{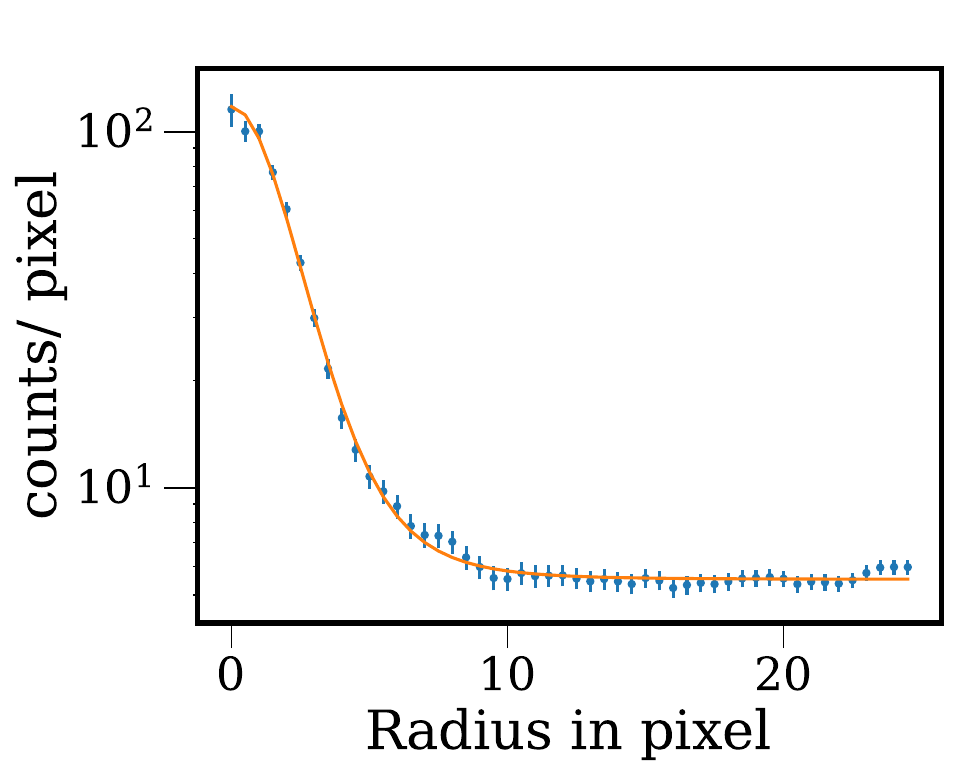}
\caption{The radial profile of HD~120190 used for PSF measurement in NUV field. The orange curve shows the fitted curve of a Moffat function added with a constant for background value.\label{fig:PSF}}
\end{figure}

Another critical parameter in source detection and photometry is the point spread function (PSF), which we determined here by using isolated bright point-like sources in the FUV and NUV images. We selected point sources in our fields  by identifying already known stars  using  the SIMBAD catalogue \citep{2000A&AS..143....9W}. We found six stars in our field that are listed in SIMBAD catalogue, only one of them was well-isolated i.e. its light profile was not contaminated significantly by other nearby sources. We selected this star, HD~120190, and generated its radial surface brightness profile by using 50 annuli with an innermost radius of 0.5 pixels and outermost radius of 25 pixels and common center at the source position determined by the {\tt SExtractor}. We then fitted the radial profile with a circular Moffat function \citep{1969A&A.....3..455M},
\begin{equation}
\centering \label{moffat}
I(r) = \frac{I_{0}}{(1+(r/\alpha)^2)^\beta}
\end{equation} 
for the source and 
a constant ($C$) to account for the background. We show the radial surface brightness profiles and the best fitting Moffat plus constant model for our NUV field image in Figure~\ref{fig:PSF}.  
We then calculated the full width at half maximum (FWHM) of the Moffat function as
\begin{equation}
\centering
\text{FWHM} = 2 \alpha \sqrt{2^{1/\beta}-1}
\label{FWHM_eq}
\end{equation}
We use the FWHM of the Moffat function  as the measure of the PSF. The calculated FWHM for our NUV and FUV images are $1.55^{\prime \prime}\pm 0.17^{\prime \prime} $ and $2.02^{\prime \prime}\pm 0.05^{\prime \prime} $, respectively.

\subsection{Signal-to-Noise Ratio (SNR) and Limiting Magnitude} \label{sect: SNR_and_limmag}
The net source counts can be written as 
\begin{equation}
    S = T - B
\end{equation}
where  $T$ is the total, source ($S$) + background ($B$), counts in the source extraction area. The variance in the net source counts is then
\begin{equation}
    \sigma_S^2 = T + \sigma_B^2 = S+B + \sigma_B^2
\end{equation}
where $\sigma_B^2$ is the variance in the total background counts. We have neglected the readout noise as UVIT uses CMOS detectors for which the readout noise is zero. If $s$ is the source count rate, $b$ is background count rate per pixel, and $\sigma_b^2$ the variance in the background count rate per pixel, then the signal-to-noise ratio (SNR) can be written as
\begin{equation}
    SNR = \frac{s\times t_{exp}}{\sqrt{s\times t_{exp} + b \times N_{pix} \times t_{exp} + N_{pix} \times \sigma_b^2 \times t_{exp}^2}}
    \label{eqn:snr}
\end{equation}
where $t_{exp}$ is the exposure time, $N_{pix}$ is the number of pixels in the source extraction region. We used the above equation to calculate the signal-to-noise ratio of our detected sources (see Section~\ref{sect:src_photometry} below).

We can also use Eqn.~\ref{eqn:snr} to derive the detectable flux for a given SNR and exposure time by solving the quadratic equation,

\begin{equation}
    s^2 - \frac{{(SNR)}^2}{t_{exp}} \times s - {(SNR)}^2  \left[\frac{b \times N_{pix}}{t_{exp}} + {\sigma_b^2 \times N_{pix}} \right] = 0
    \label{eqn:lim_mag}
\end{equation}
and taking the positive root.

The magnitude of a source in AB system can be measured using the equation
\begin{equation}
\centering \label{mag_eqn}
 m_{AB} = - 2.5 \log(CPS) + \text{zero-point magnitude}
\end{equation}
 where CPS is the source count rate ($s$) in counts per second and the zero-point magnitudes are the the values measured by  \cite{2017AJ....154..128T} for each of the filters. For our NUV filter zero-point magnitude is $18.50 \pm 0.07$ and for our FUV filter it is $17.78 \pm 0.01$. Using the limiting flux value from Eqn.~\ref{eqn:lim_mag}, we then obtained the value of limiting magnitude from Eqn.~\ref{mag_eqn}.
 
\begin{figure*}[ht!]
\plotone{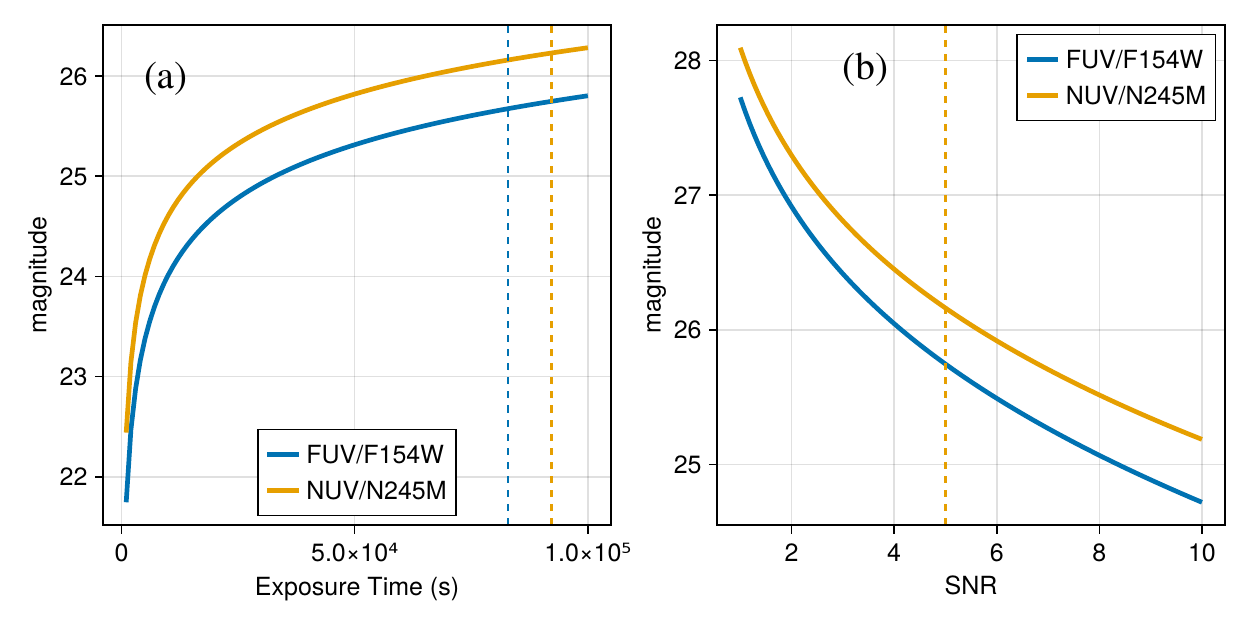}
\caption{Limiting magnitudes in the FUV/F154W and NUV/N245M bands as a function of (a) exposure time for an SNR of 5, and (b) signal-to-noise ratio for the actual FUV and NUV exposure times. The dashed lines in (a) represent the net FUV and NUV exposure times for the IC~4329A field. Our choice of an SNR of 5 for the source detection is marked as a dotted line in (b).
\label{limmag}}
\end{figure*}

For a point source (considering a circular profile of $10\arcsec$ radius) with SNR of 5, we found a limiting magnitude of 26.15 in the NUV band and 25.75 in FUV band. In Figure~\ref{limmag}(a), we show the variation in the limiting magnitude with exposure time as calculated using Eqn.~\ref{eqn:lim_mag} and \ref{mag_eqn} for an SNR of 5. The vertical dashed lines mark the net exosures of our FUV and NUV observations. Figure~\ref{limmag}(b)  depicts the variation in limiting magnitude with the SNR for the net exposures of our observations.  Following these equations we have measured SNR and AB magnitudes of our detected sources for both the fields, as mentioned in the Section \ref{sect:src_photometry}.

\subsection{Source Detection and photometry} \label{sect:src_photometry}

\begin{figure*}[ht!]
\gridline{\fig{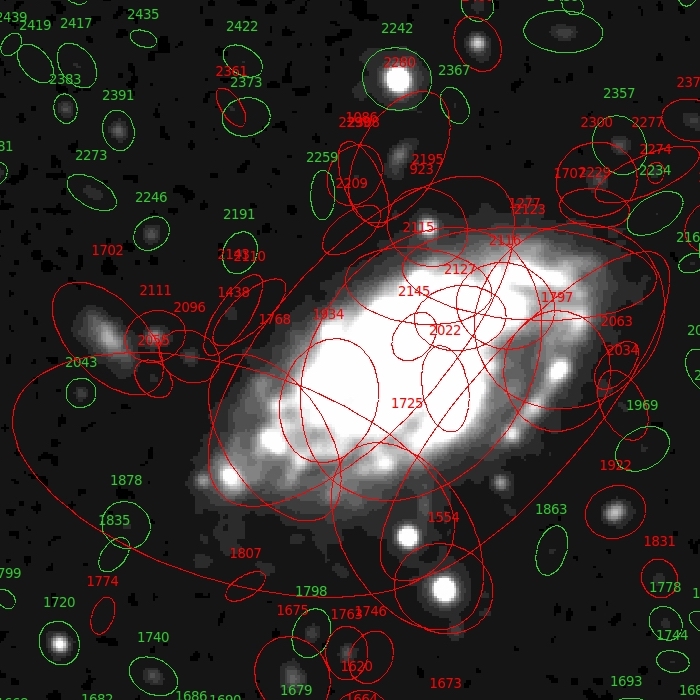}{0.5\textwidth}{a}
            \fig{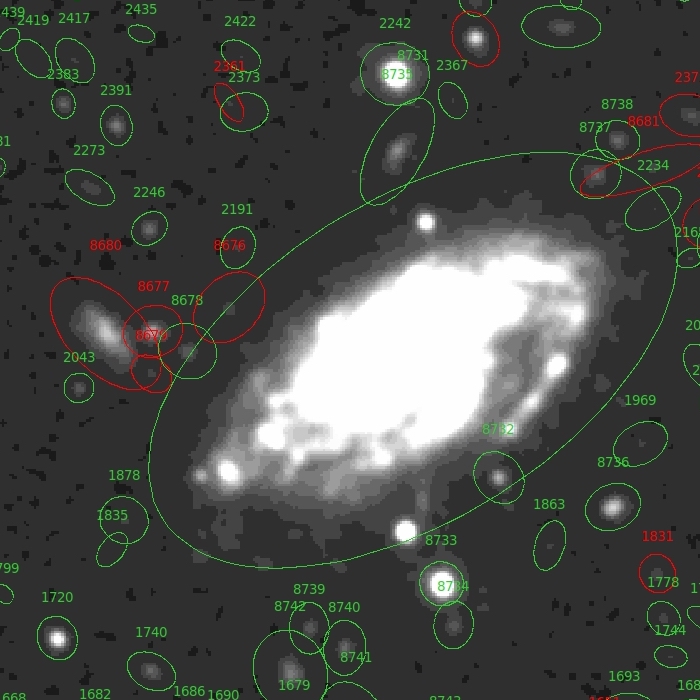}{0.5\textwidth}{b}
            }
\caption{Detection of an extended source, a galaxy in this case, in the NUV band image using two different sets of detection criteria. The green and red ellipses in both the images indicate the extraction regions of the detected sources while the source id in the corresponding catalogue file is mentioned next to each source region. left Panel: initial detection of the galaxy by {\tt SExtractor} as a group of several closely spaced sources due to its clumpy structure in UV band when we used the detection parameters of the main configuration file as listed in Table~\ref{table:main config param table}. Right Panel: Proper detection of the galaxy as a single extended source when the detection parameters listed in aux. catalogue 5 column in  Table~\ref{table:aux_config_file_NUV} is used instead.
\label{fig:detection_process}}
\end{figure*}

We used the final merged images in the FUV and NUV bands for source detection and photometry. First, we modified those images by subtracting the corresponding mean background from all the pixels inside our common elliptical field (see Figure~\ref{fig:field}) and making the counts in all the pixels outside the common field zero. To detect the sources and to measure their photometric properties we ran {\tt SExtractor} on these background-subtracted images. We list the {\tt SExtractor} parameters in Table \ref{table:main config param table} that we used in this stage to detect sources in the FUV and NUV bands.
By examining image segments and the Kron profile of extracted sources, we noticed that the {\tt SExtractor} tool provided fairly accurate results for point-like sources but in the case of extended sources, the results were not satisfactory as it detected multiple sources within the extension of a diffuse source with spatial structures which are common in galaxies.  This is  not unexpected as the {\tt SExtractor} tool was  developed primarily to detect sources in the optical band. Galaxies generally show  smooth surface brightness in the optical band while spatial structures  primarily due to star forming regions are prominent in the UV band. 
To overcome this difficulty with the detection of extended sources, we used separate configuration files with different values of relevant detection parameters that allowed us to extract the extended sources and the sources around them, thus generating  auxiliary catalogue files. We had to use more than one such auxiliary configuration files to detect the extended sources located in different regions in our fields. We then manually replaced the sources detected in the original catalogue with the properly detected sources in auxiliary catalogues. In the Tables~\ref{table:aux_config_file_NUV} and \ref{table:aux_config_file_FUV} we have mentioned the different detection criteria we used to generate those auxiliary catalogues and the number of sources that were replaced using the corresponding catalogue file. In Figure~\ref{fig:detection_process}, we show an extended source -- a galaxy in one part of the NUV field. The left panel in the figure shows how {\tt SExtractor} (with the detection parameters  listed in Table~\ref{table:main config param table}) originally detected the extended source as a group of several closely spaced sources possibly due to the clumpy structure within the extended galaxy. The image in the right panel of Figure~\ref{fig:detection_process}  shows the identification of extended emission as one source when we changed our detection criteria as listed in aux. catalogue 5 column in  Table~\ref{table:aux_config_file_NUV}.

\begin{deluxetable*}{ccc}
\tablecaption{Details of detection parameters of the  {\tt SExtractor} tool used for the main catalogue file for NUV and FUV field.  \label{table:main config param table}}
\tablewidth{0pt}
\tablehead{\colhead{Parameter} & \colhead{main catalogue file for NUV}  & \colhead{Main catalogue file for FUV}}
\startdata
DETECT\_THRESH & 2.0 & 1.5\\
DETECT\_MINAREA & 10 & 10\\
DEBLEND\_NTHRESH & 32 & 32\\
DEBLEND\_MINCOUNT & 0.001 & 0.05\\
Number of detected sources & 8356 & 1611\\
Number of sources taken & 7226 & 1529\\
\enddata
\end{deluxetable*}

\begin{deluxetable*}{lccccccc}
\tablecaption{Details of detection parameter values used in different auxiliary catalogue files for {\tt SExtractor} runs in case of NUV field.\label{table:aux_config_file_NUV}}
\tablewidth{0pt}
\tablehead{\colhead{Parameter} & \colhead{Aux. Cat. 1} & \colhead{Aux. Cat. 2}  & \colhead{Aux. Cat. 3} & \colhead{Aux. Cat. 4} & \colhead{Aux. Cat. 5} & \colhead{Aux. Cat. 6} & \colhead{Aux. Cat. 7}
}
\startdata
DETECT\_THRESH & 3.0 & 6.0 & 2.0 & 5.0 & 5.0 & 3.0 & 10.0 \\
DETECT\_MINAREA & 10 & 20 & 10 & 20 & 4 & 15 & 20 \\
DEBLEND\_NTHRESH & 10 & 32 & 32 & 10 & 10 & 10 & 10 \\
DEBLEND\_MINCOUNT & 0.1 & 0.1 & 0.01 & 1.0 & 1.0 & 1.0 & 1.0 \\
Number of Sources & 444 & 38 & 34 & 14 & 19 & 13 & 4 \\
\enddata
\end{deluxetable*}

\begin{deluxetable*}{cccc}
\tablecaption{Details of detection parameter values used in different auxiliary catalogue files for {\tt SExtractor} runs in case of FUV field.\label{table:aux_config_file_FUV}}
\tablewidth{0pt}
\tablehead{\colhead{Parameter} & \colhead{Aux. Cat. 1} & \colhead{Aux. Cat. 2}  & \colhead{Aux. Cat. 3}}
\startdata
DETECT\_THRESH & 2.0 & 3.5 & 2.0 \\
DETECT\_MINAREA & 8 & 10 & 25 \\
DEBLEND\_NTHRESH & 32 & 32 & 5 \\
DEBLEND\_MINCOUNT & 1.0 & 1.0 & 1.0\\
Number of sources & 14 & 5 & 1\\
\enddata
\end{deluxetable*}

\begin{figure*}
 \gridline{\fig{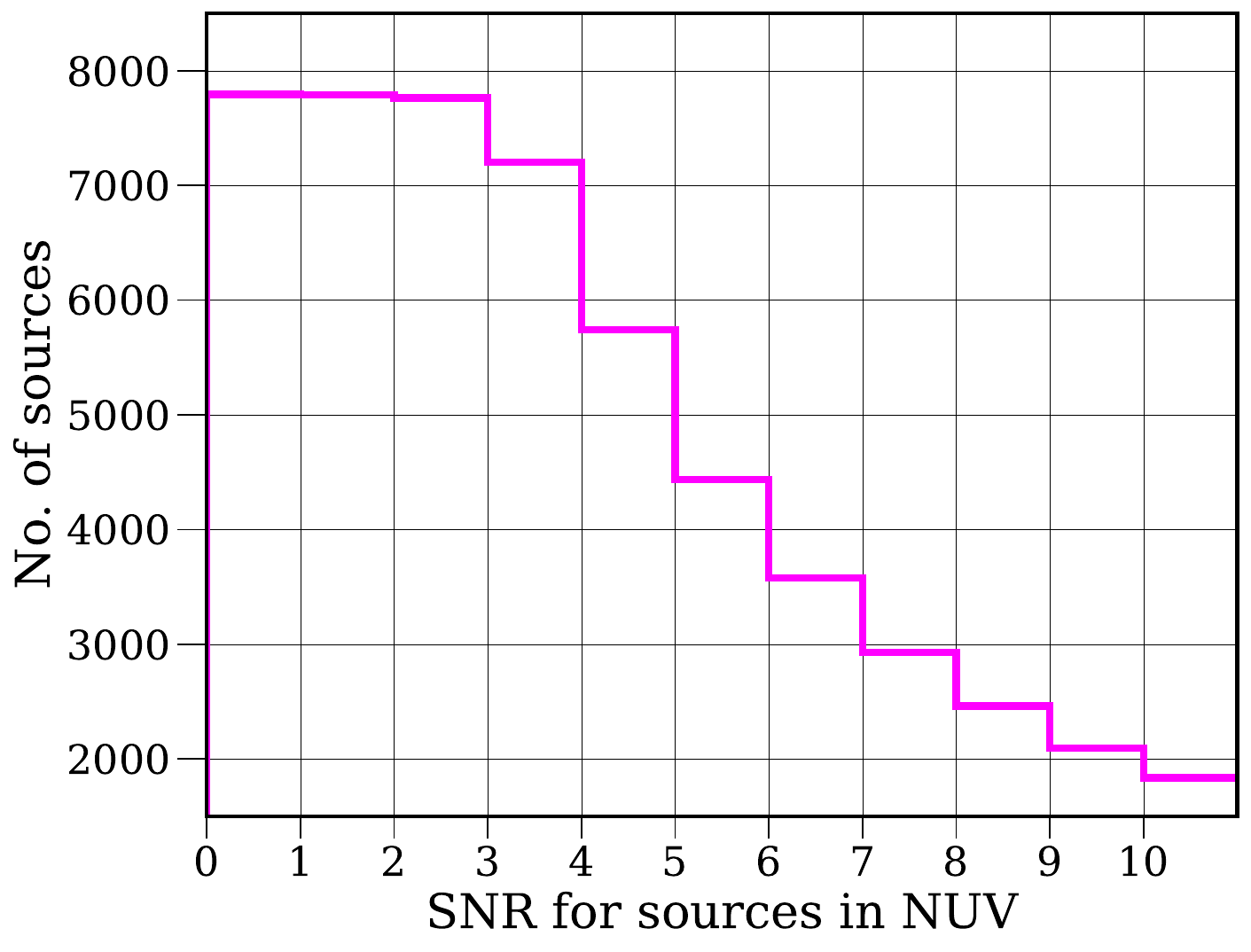}{0.5\textwidth}{}
            \fig{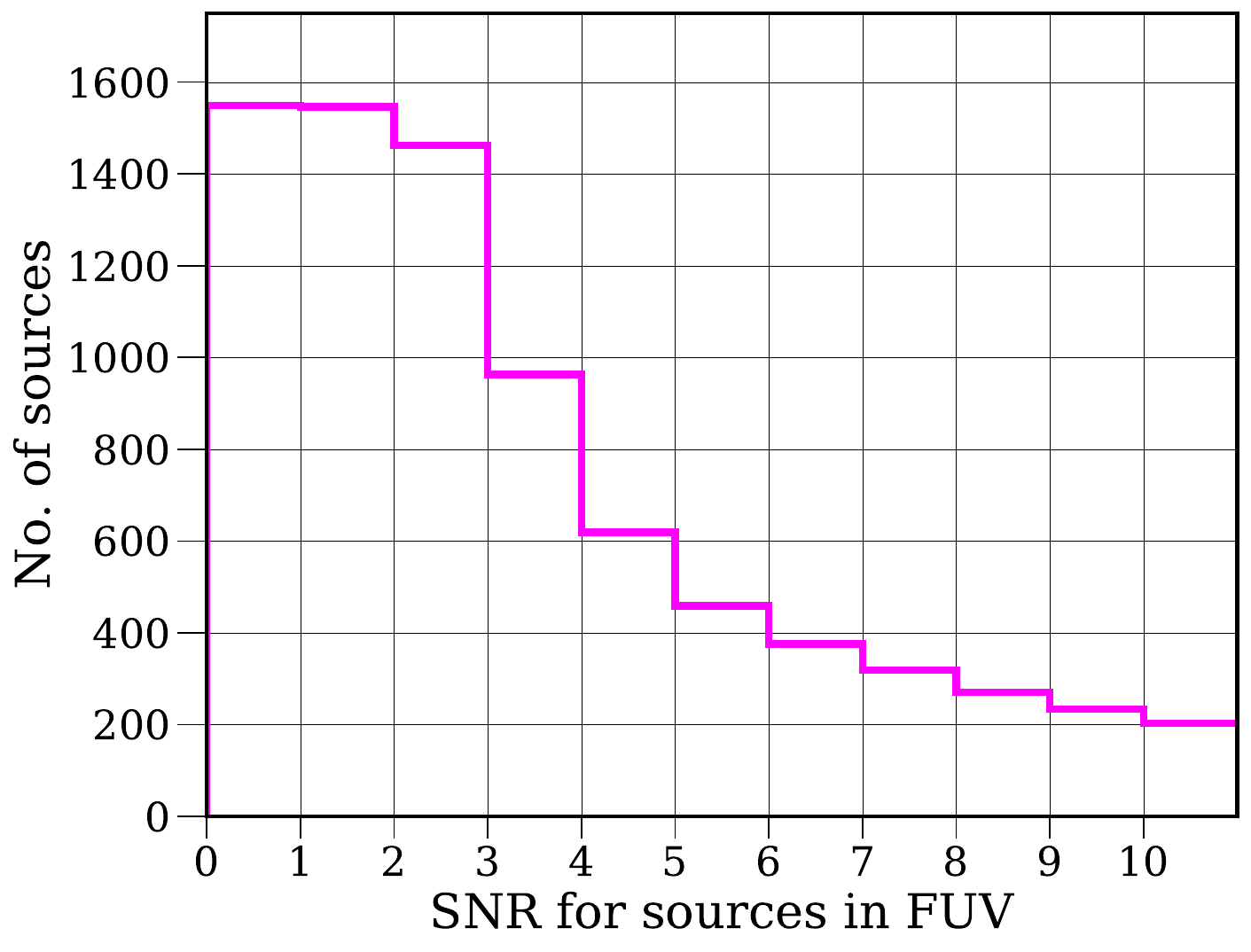}{0.5\textwidth}{}
            }
\caption{Cummulative histogram of calculated SNR of all the sources in the NUV (left panel) and the FUV (right panel) field.
\label{fig:cummulative_snr}}
\end{figure*}

After detecting all the sources present in our fields (for both FUV and NUV fields) we calculated the SNR and AB magnitudes for all those sources using the equations \ref{eqn:snr} and \ref{mag_eqn} mentioned in Section \ref{sect: SNR_and_limmag}. Figure~\ref{fig:cummulative_snr} shows cumulative histograms indicating total number of sources above a certain SNR  as marked on the x-axis for both the NUV and FUV fields. For the NUV field we found a total of 4437 sources above an SNR of 5 with the faintest source having magnitude 26.0 and in case of FUV field we found a total of 457 sources above an SNR of 5 with the faintest source having magnitude 25.1.

\begin{figure*}
\gridline{\fig{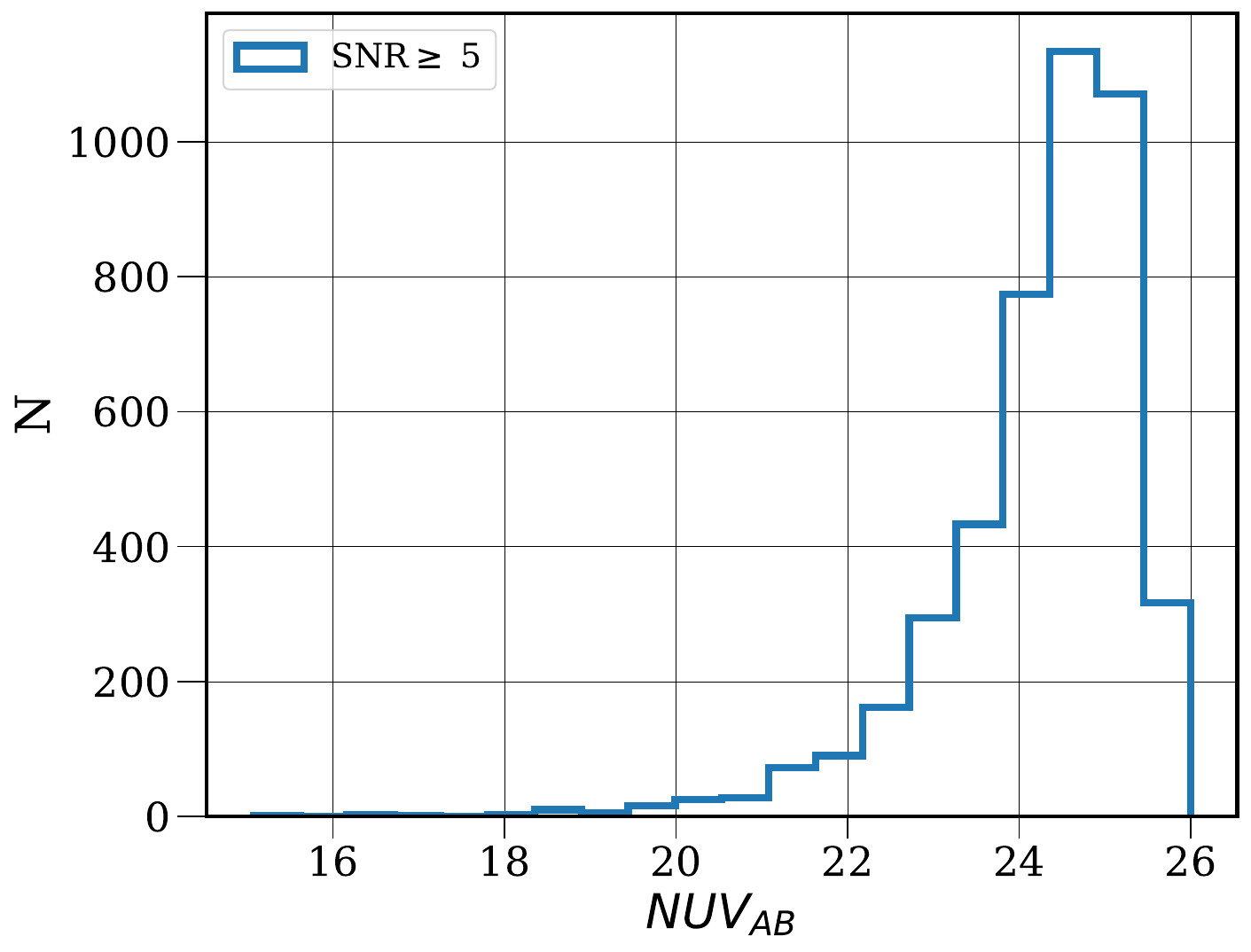}{0.5\textwidth}{}
    \fig{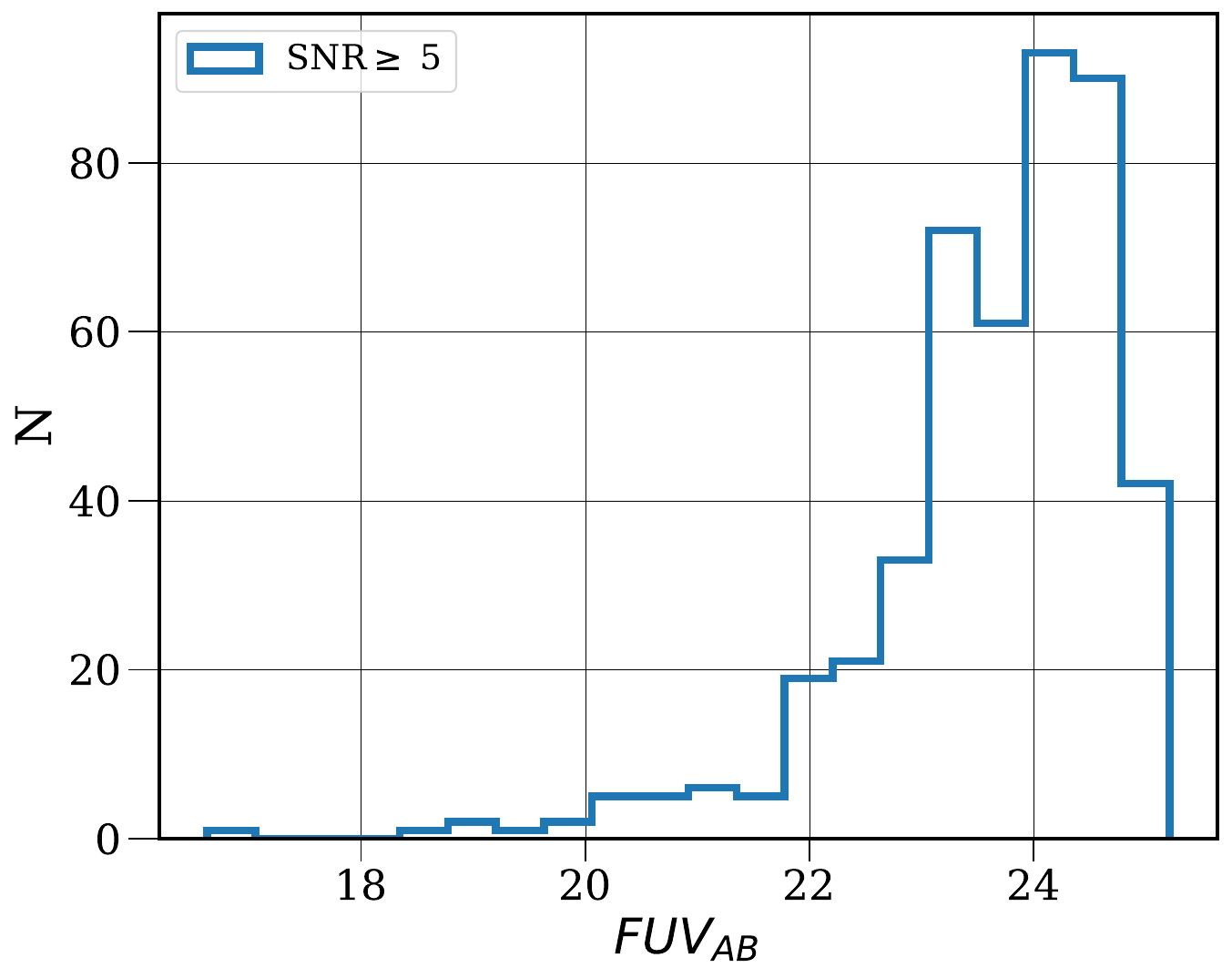}{0.5\textwidth}{}
    }
\caption{The histogram of AB  magnitude of all the sources in the NUV (left panel) and the FUV (right panel) field for sources above SNR of 5.
\label{fig:snr_mag_diagram}}
\end{figure*}

\subsection{Aperture Correction}
\begin{figure*}
\gridline{
    \fig{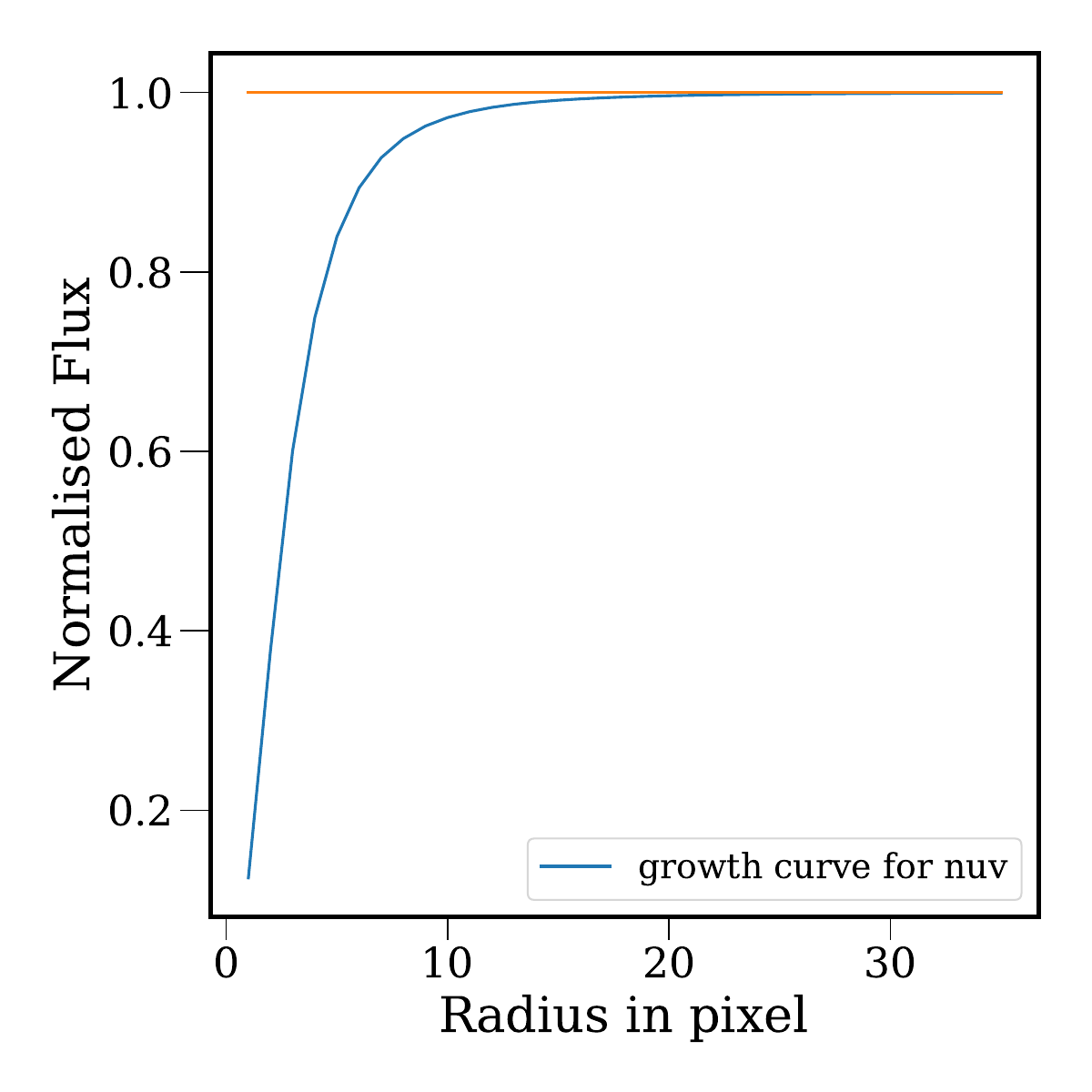}{0.45\textwidth}{}
    \fig {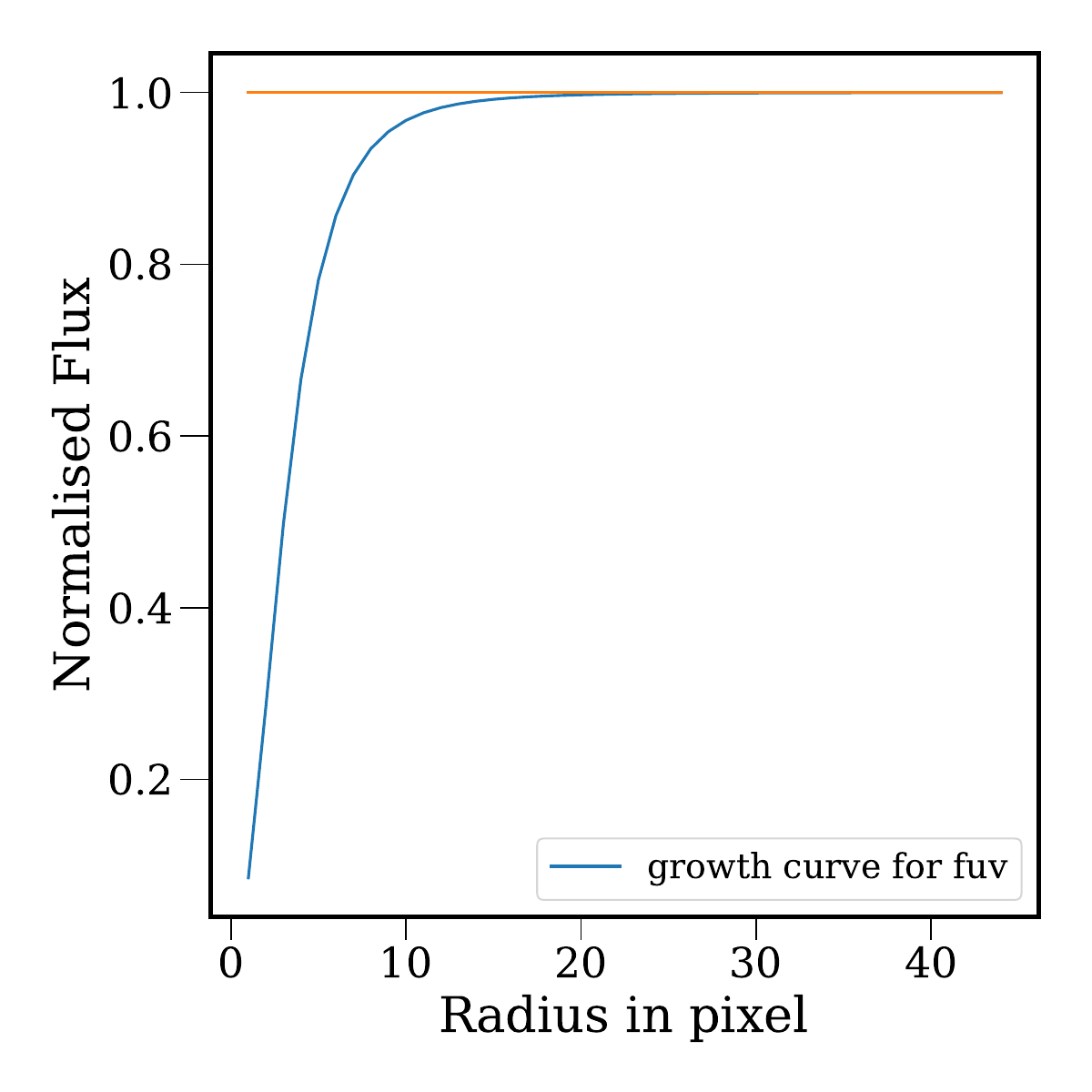}{0.45\textwidth}{}
}
\caption{The growth curves (blue) for the NUV (left panel) and FUV (right panel) bands, showing  the fractional enclosed flux as a function of radius in pixels. Each pixel measures $0.416^{\prime \prime}$.
\label{fig:growth_curves}}
\end{figure*}

For photometric measurements using SExtractor, we used the Kron aperture corresponding to a Kron factor of 2.5 for each source in our catalogue. Since the Kron radius for an individual source may depend upon the position of the neighbouring sources and the required de-blending, the flux measured for a point source by the SExtractor may not represent the true flux of the source. The extraction region or the aperture size used for a point source may be smaller than the full extend of the PSF, and  may not include the entire flux. Since the PSF of the FUV and NUV channels extend upto $\sim 10{\rm~arcsec}$, it is important to account for any loss due to smaller aperture size. We therefore performed aperture correction to the SExtractor-measured flux for point sources. For this purpose, we derived growth curve from the PSF for the NUV and FUV bands. We show the growth curves in Figure~\ref{fig:growth_curves}. These curves show that the wings of the PSF extend upto $\sim 25$ pixel radius for both the NUV and the FUV bands. From these growth curves we calculated the  percentage of total source flux measured in a circular extraction region of varying radius and it is given in Table \ref{Table:flux_percentage}. Using this table we corrected the measured flux as well as the magnitude for each point source in our catalogue. To apply this aperture correction in case of an elliptical source extraction region we found an equivalent radius of a circle with same area as the ellipse. 

\begin{deluxetable*}{ccc}
\tablecaption{Enclosed flux (in \%) as a function of radius in pixels for the FUV and NUV bands.\label{Table:flux_percentage}}
\
\tablewidth{0pt}
\tablehead{\colhead{Radius (pixels)} & \colhead{NUV flux (\%)} & \colhead{FUV flux (\%)}}
\startdata
1 & 12.48 & 8.08 \\ 
2 & 37.87 & 26.99 \\
3 & 60.12 & 47.31 \\
4 & 74.92 & 63.69 \\
5 & 83.94 & 75.26 \\
6 & 89.38 & 83.01 \\
7 & 92.73 & 88.13 \\
8 & 94.86 & 91.53 \\
9 & 96.26 & 93.83 \\
10 & 97.21 & 95.41 \\
11 & 97.87 & 96.52 \\
12 & 98.34 & 97.31 \\
13 & 98.68 & 97.89 \\
14 & 98.94 & 98.32 \\
15 & 99.14 & 98.65 \\
16 & 99.29 & 98.90 \\
17 & 99.41 & 99.09 \\
18 & 99.50 & 99.24 \\
19 & 99.58 & 99.37 \\
20 & 99.64 & 99.46 \\
21 & 99.69 & 99.54 \\
22 & 99.73 & 99.61 \\
23 & 99.76 & 99.66 \\
24 & 99.79 & 99.71 \\
25 & 99.82 & 99.74 \\
26 & 99.84 & 99.77 \\
27 & 99.85 & 99.80 \\
28 & 99.87 & 99.82 \\
\enddata
\end{deluxetable*}

\subsection{Cross-matching between FUV and NUV catalogues}
After detecting all the sources in both the fields, we tried to cross-match our FUV catalogue with our NUV catalog to see whether all FUV sources have their counterparts in the NUV band. We first found the nearest NUV source for each FUV source in our catalogue. 
We considered the closest NUV source to a particular FUV source as its counterpart if ($i$) the nearest NUV source lies within the $99.7\%$ astrometric confidence circle around the FUV source, or ($ii$) the nearest NUV source does not fulfill the first criterion but the center of the NUV source lies within the Kron aperture of the FUV source.
As the spatial distribution of an extended source in the FUV and NUV bands can be different,  the astrometric position of such a source measured based on its light profile can be slightly shifted in the two bands. In such a case, the second criteria helps us to determine the counterparts even when  the first criteria is not satisfied.

\begin{figure*}
\gridline{\fig{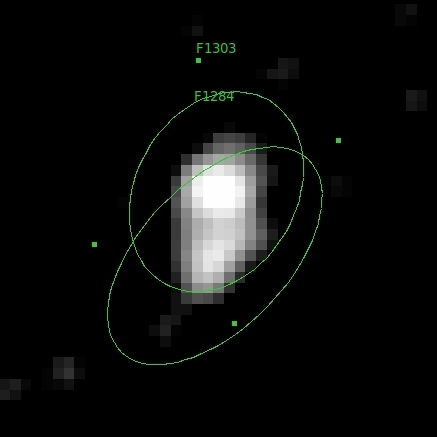}{0.5\textwidth}{a}
            \fig{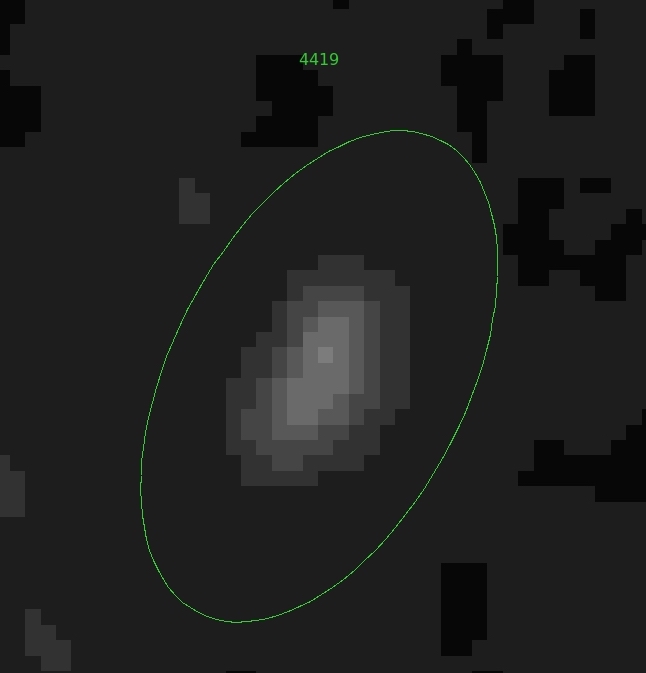}{0.5\textwidth}{b}
            }
\caption{Left Panel image shows the two FUV sources that has  both been identified with a single NUV source shown in the right panel image. The green ellipse shows the source region and the number next to it is the ID of that source in that particular (FUV or NUV) catalogue. In the FUV field image we can see two sources with overlapping light profile but with distinguised core regions indicating presence of two sources although in NUV image these two sources can not be identified separately.
\label{fig:NF_counterpart_src}}
\end{figure*}

Comparing the catalogues of sources above SNR of 5, using the first and second criteria we found 452 and 5 FUV sources with NUV counterparts, respectively. All FUV sources were uniquely assigned to a distinguished NUV source, except for only two FUV sources that were both assigned to a single NUV source.
These two FUV sources and their assigned counterpart is shown in Figure~\ref{fig:NF_counterpart_src} where the green regions show the extraction region of each source and the ID of that source (both in NUV and FUV catalogue) is shown near the region. The properties of these two FUV sources are listed in table \ref{table:counterpart_src_FUV}. In our final catalogue we have assigned the FUV source with ID F1303 (as shown in the left panel image of Fig \ref{fig:NF_counterpart_src}) as the FUV counterpart of the NUV source shown in the right panel of the figure.

\begin{deluxetable*}{ccccc}
\tablecaption{FUV Properties of the two sources that has been identified with same NUV counterpart \label{table:counterpart_src_FUV}}
\tablewidth{0pt}
\tablehead{\colhead{FUV ID} & \colhead{Flux} & \colhead{Error on flux} & \colhead{AB magnitude} & \colhead{Error on AB magnitude}\\
 & (ergs ${\rm{cm}}^{-2} {\rm{sec}}^{-1}$ \AA${}^{-1}$) & (ergs ${\rm{cm}}^{-2} {\rm{sec}}^{-1}$ \AA${}^{-1}$) & &  }
\startdata
1303 & 1.03e-17 & 1.07e-18 & 24.12 & 0.11\\
1284 & 8.30e-18 & 1.09e-18 & 24.36 & 0.14\\
\enddata
\end{deluxetable*}

\begin{figure}
\plotone{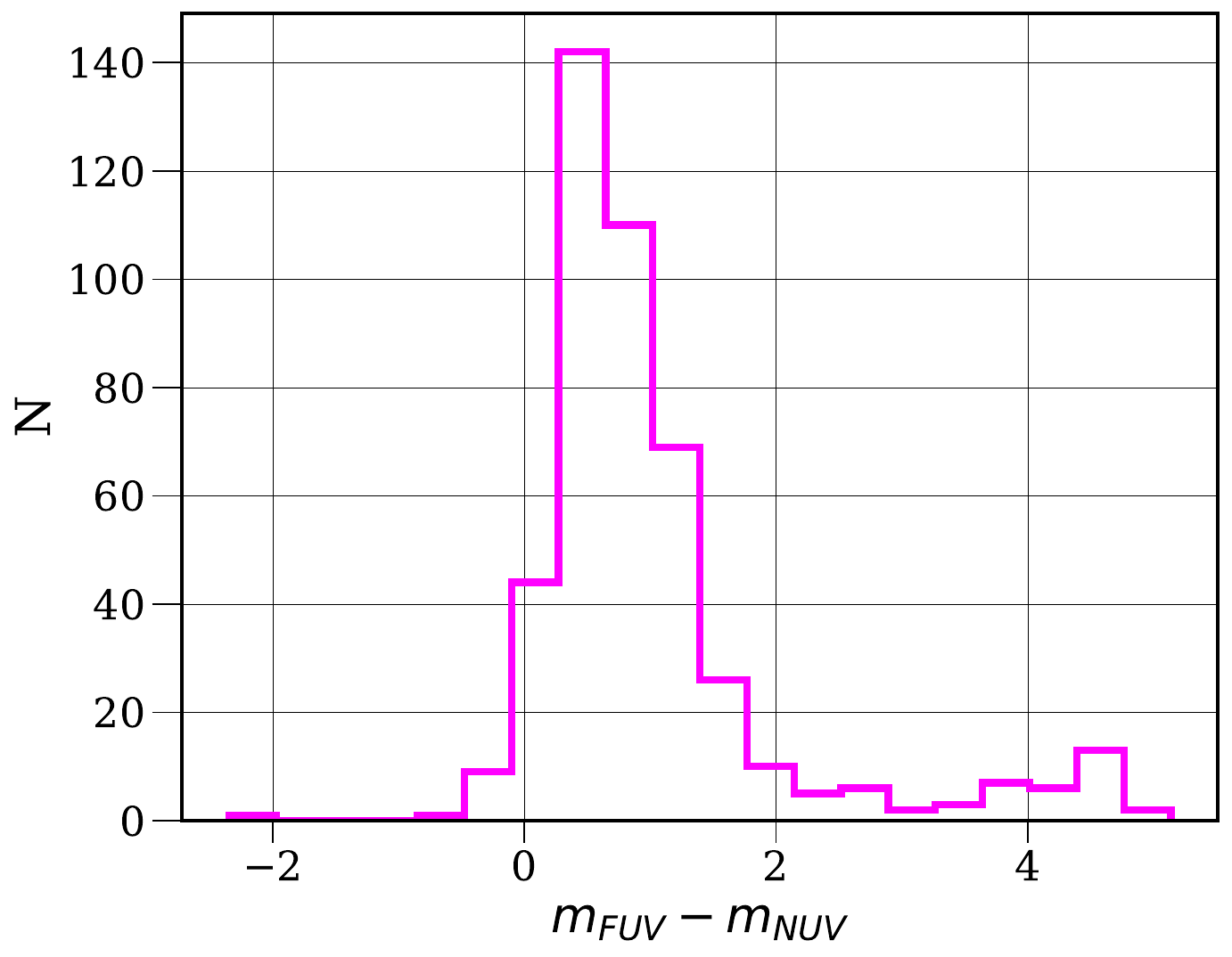}
\caption{Histogram of UV color ($m_{FUV} - m_{NUV}$) for all 456 sources in our catalog that are detected both in the NUV and FUV bands above the SNR of 5. A total of 17 sources are found with negative color values i.e. they are brighter in the FUV than in the NUV band \label{fig:NF_color}}
\end{figure}

We have also calculated $m_{FUV} - m_{NUV}$ for all 456 sources that have been detected both in NUV and FUV. We  show a histogram of the color of these sources in Figure~\ref{fig:NF_color}. We found 17 sources in total that are brighter in the FUV than in the NUV band.

\subsection{Optical, X-ray and Radio counterparts}\label{sect:multiband_counterparts}
 We have cross-matched our UVIT catalogue with the \gaia{} DR3 catalogue \citep{2022arXiv220800211G} for optical counterparts. The criterion we have used to identify the optical counterpart in the Gaia catalogue is as follows. For a given source in our UVIT catalogue, if the nearest \gaia{} source position lies within $99.7\%$ confidence level for astrometric accuracy of corresponding field image, we considered the \gaia{} source as the optical counterpart of the UVIT source. Since astrometric accuracy of Gaia sources are much better than the UVIT astrometric accuracy by orders of magnitude we did not consider the astrometric error of Gaia sources. Following this criterion, we found 651 cross-matched \gaia{} sources in the NUV field out of which 50 sources have FUV counterparts also.

 We also cross-matched our catalogue with the \xmm{} serendipitous catalog  from the 12th data release (DR12) \citep{2020A&A...641A.136W} to find the X-ray counterparts. The DR12 catalogue lists the sources that have been observed with \xmm{} in the 0.2-12 KeV band. For the cross-matching we used the same criterion i.e., angular separation between an UVIT source and X-ray counterpart is within the $99.7\%$ confidence level of astrometric accuracy. In this case we have also taken into account the astrometric  accuracy of \xmm{} sources since unlike high accuracy of \gaia{} sources astrometry, \xmm{} sources are comparable in magnitude with the astrometric accuracy of our sources. We found X-ray counterparts for 97 NUV sources and out of which 25 sources have FUV counterparts also. In the final catalogue file we have mentioned the \gaia{} ID, \gaia{} position, \gaia{} flux and error on that flux, \gaia{} proper motion and \gaia{} classification according to their proper motion for the Optical counterparts of our sources. We also mention the \gaia{} G band magnitudes of the optical counterparts in AB magnitude system which we calculated from \gaia{} flux and zero-point magnitude of 25.8010\footnote{available at \url{https://www.cosmos.esa.int/web/gaia/edr3-passbands}}. For the X-ray counterparts of our sources we have mentioned their XMM ID, XMM position , X-ray flux and error on that flux in our final catalogue.

 We searched for radio counterparts to our UV sources by cross-matching our catalogue with the NRAO VLA Sky Survey (NVSS) catalogue \citep{Condon_1998} that lists all the radio sources north of $\delta (J2000) = -40^{\circ}$ and with a  radio flux density $ S > 2.5$ mJy at 1.4 GHz. We found only 7 radio sources in our field. 
 To find the UV counterparts for these sources, we used circular regions around the radio sources with the radio positional errors as the radii and treated a radio source as the counterpart if a UV source position lies within the circular region. We only considered the radio positional errors as they are an order of magnitude larger than the UV positional errors. Thus we found 4 radio sources that have UV counterparts in our catalogue. Two of these sources are identified as Seyfert galaxy IC~4329a (source id 4194 in our catalogue) and optical emmision line galaxy IC~4327 (source id 4255 in our catalogue), respectively. The third radio source has two possible UV counterparts -- LEDA 719417 (UV source id 3909) and LEDA 719364 (UV source id 3915) and these two sources seems to be in an interacting system of galaxies. All the details of the radio counterparts of these sources are mentioned in Table~\ref{table:radio_counterpart}. We have also discussed detailed UV morphology of IC~4327 and the seemingly interacting system of LEDA~719417 and LEDA~719364 in Section \ref{sect:int_sources}.

\begin{deluxetable*}{cccccccc}
\tablecaption{Properties of the sources identified with radio counterparts \label{table:radio_counterpart}}
\tablewidth{0pt}
\tablehead{\colhead{Source ID} & \colhead{NVSS ID} & \colhead{RA J2000} & \colhead{Error on RA} & \colhead{DEC J2000} & \colhead{Error on DEC} & \colhead{$S_{1.4 GHz}$} & \colhead{Error on $S_{1.4 GHz}$} \\
\colhead{(UV catalogue)} & & \colhead{(NVSS)} & \colhead{(NVSS)} & \colhead{(NVSS)} & \colhead{(NVSS)} &  & \\
 & & \colhead{(degree)} & \colhead{(s)} & \colhead{(degree)} & \colhead{(arcsec)} & \colhead{(mJy)} & \colhead{(mJy)}}
\startdata
4194 & 134919-301833 & 207.330125 & 0.0 & -30.309361 & 0.6 & 66.4 & 2.0 \\
4255 & 134843-301259 & 207.183125 & 0.5 & -30.216583 & 9.7 & 2.9 & 0.6 \\
3909 & 134917-301543 & 207.323167 & 0.5 & -30.262000 & 12.2 & 2.9 & 0.6 \\
3915 & 134917-301543 & 207.323167 & 0.5 & -30.262000 & 12.2 & 2.9 & 0.6 \\
4302 & 134915-300609 & 207.314000 & 0.0 & -30.102500 & 0.6 & 46.6 & 1.5 \\
\enddata
\end{deluxetable*}

 \subsection{Categorization of the UVIT sources}
 
To distinguish between point sources and extended sources in our field, we  used the {\tt SExtractor} parameter named `CLASS\_STAR'. This parameter indicates the probability of a source being a point source or an extended source. A `CLASS\_STAR' value close to 1 indicates the source to be a point source and a value close to 0 suggests spatially extended source \citep{1996A&AS..117..393B}.  To compute this parameter {\tt SExtractor} uses a neural network that uses the parameter `SEEING\_FWHM'  specified by the user \citep{1996A&AS..117..393B}. We  used measured FWHM of the point spread function for our NUV image as the `SEEING\_FWHM' parameter value in {\tt SExtractor} configuration file.

\begin{figure}
\plotone{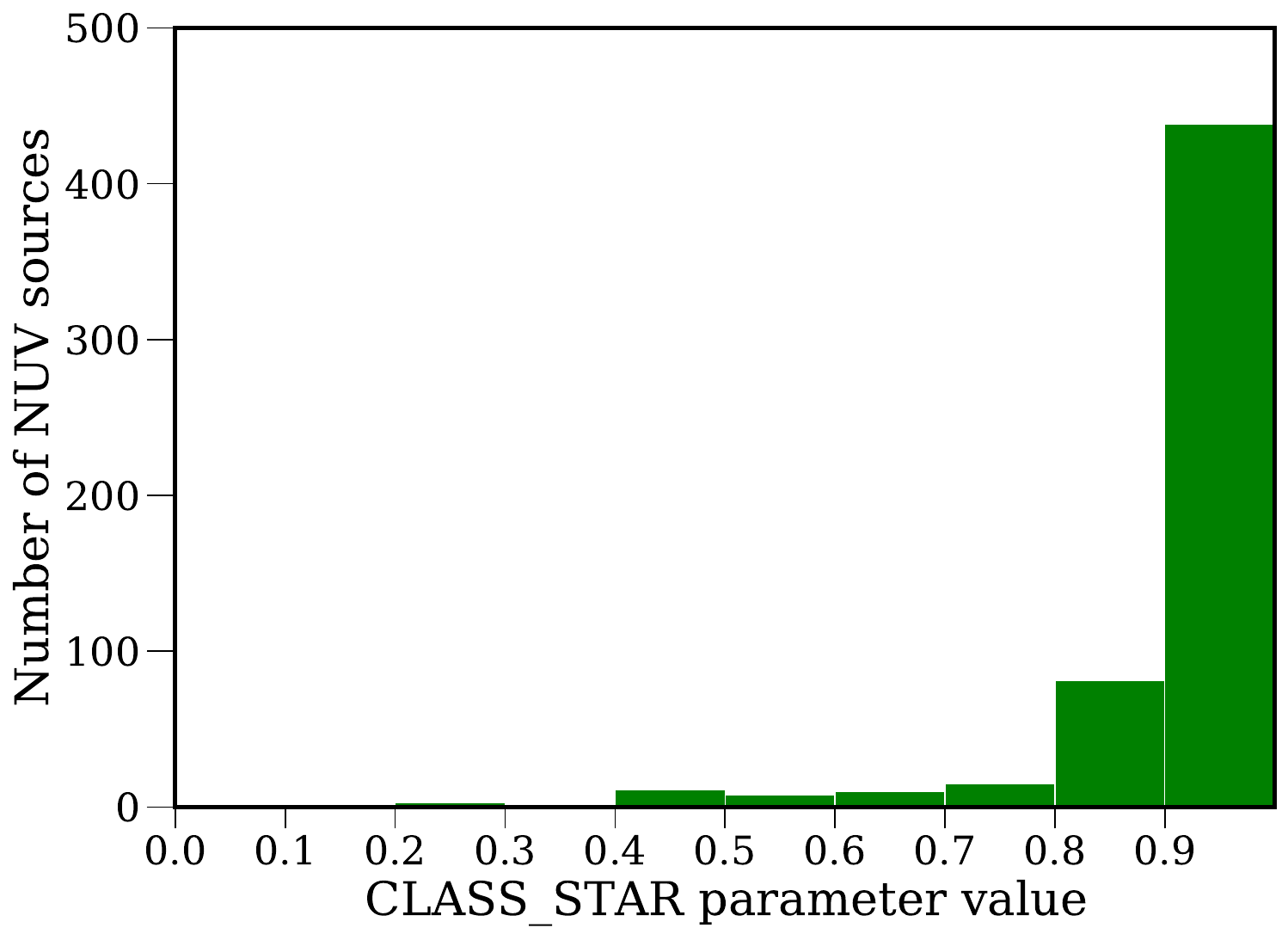}
\caption{The histogram of the CLASS\_STAR value given by {\tt SExtractor} for all the \gaia{} stars present in our catalogue. The \gaia{} sources for which the measured value of any of the two components of proper motion is significant above $3\sigma$ level are considered to be stars.\label{fig:class_star_dist}}
\end{figure}

Since {\tt SExtractor} parameter `CLASS\_STAR' is only an indication to distinguish between a point-like and an extended source, we have tested the `CLASS\_STAR' value against known Gaia sources. Among  651 sources above the SNR of 5, we identified point-like sources that had a non-zero proper motion (significant above $3 \sigma $level) which are clearly the stars. Thus we found 568 stars above SNR of 5, and we checked the `CLASS\_STAR' values for those sources. Figure~\ref{fig:class_star_dist} shows a histogram for the `CLASS\_STAR' variable  for the 568 Gaia stars above SNR of 5. We found that $96\%$ of the stars (544 out of 568) detected above SNR of 5 have the CLASS\_STAR value greater than 0.6. Therefore, we have classified the UVIT sources with CLASS\_STAR value greater than 0.6 as point sources with $96\%$ confidence. Following this categorization we have found a total of 1872 point sources and 2566 extended sources in our catalogue.

Although currently we do not have spectroscopic data to find the redshift of the sources in our catalogue our field harbours some of the sources of Shapley supercluster, also known as IC 4329 cluster. So, we cross-correlated our catalogue with redshift database towards Shapley supercluster region  presented by \citep{2020A&A...638A..27Q} and identified total 17 sources in our field for which redshift was available. We have added the redshift informations for these 17 sources in our catalogue.

\section{Results \& Discussions}\label{sect:results}
In this study, we analysed the data from the UVIT field around the Seyfert galaxy IC4329a and produced a catalogue of 4437 sources. We classified these sources as point-like or extended, and searched for optical and X-ray counterparts. We investigated the nature of point-like sources with \gaia{} counterparts without appreciable proper motion. We also characterised the variability of the UV sources. Below we describe our catalogue and source characterisation through color-color diagram and UV variability.

\subsection{Final source catalogue}
Our final source catalogue contains the position and photometric properties of 4437 NUV sources detected above a SNR  of 5. We have  categorized our sources as a probable point or extended sources with $95\%$ confidence. Our final catalogue contains 1871 point sources and 2566 extended sources. For extended sources we have provided the value of semi-major axis and semi-minor axis of the Elliptical Kron aperture (corresponding to Kron factor 2.5) for the source which has been used for all the photometric measurements. We detected sources up to a magnitude of 26.0 in the NUV band, 25.1 in the FUV band. 
Out of these 4437 NUV sources, only 456 sources are detected in the FUV band, 651 sources have  \gaia{}/optical and 97 sources have \xmm{} X-ray counterparts. We found only 10 sources with counterparts in all the three bands i.e. FUV, optical and X-ray. In Table~\ref{table:catalog_param}, we provide a detailed description of the columns in our catalogue. We are making available the catalogue in FITS format.

\begin{table*}
\caption{Description of the source properties mentioned in final catalogue.\footnote{For each source all the columns related to FUV, \gaia{} or \xmm band is set to zero if their is no counterpart of the source in the respective band. Only for the column `UV color' we have kept the value as `-99.99' if there is no FUV counterpart for a source}}
\footnotesize
\begin{tabular}{m{3cm}  m{12cm}}
\hline 
Column Name & Description \\
\hline
ID &   Source identification number in the catalogue \\
NUV\_RA\_J2000 &  Right ascension in degrees (in the NUV band)\\
NUV\_DEC\_J2000 &  Declination in degrees (in the NUV band)  \\
A\_NUV & Semi-major axis of the elliptical Kron aperture\footnote{Corresponding to Kron factor 2.5 \label{footnote_1}} in arcsec (in NUV band)\footnote{Given only for the extended sources. For point sources value is quoted as `0.0' \label{footnote_2}} \\
B\_NUV & semi-minor axis of the elliptical Kron aperture\footref{footnote_1} for the source in arcsec (in NUV band)\footref{footnote_2}\\
THETA\_NUV & Angle for the elliptical Kron aperture\footref{footnote_1} in degrees (in NUV band)\footref{footnote_2}\\
NUV\_FLUX   &   Estimated flux of the source in the NUV band in units of  ergs ${\rm{cm}}^{-2} {\rm{sec}}^{-1}$ \AA${}^{-1}$ \\
NUV\_FLUX\_ERR    &  Error on NUV\_FLUX  \\
NUV\_MAG    &  Estimated AB magnitide  of the source in the NUV band\\
NUV\_MAG\_ERR &   Error on NUV\_MAG \\
NUV\_FLAG & Extraction Flag for the source in the NUV band\footnote{A value of `1' means the photometric measurements are affected due to nearby bright sources (more than 10\% area affected), otherwise the flag value is `0'}\\
SRC\_CATEGORY & Category of the source\footnote{`1' : point source and `0' : extended source \label{footnote_3}} \\
NUV\_SNR & Signal to Noise ratio of the source in the NUV band \\
NUV\_VAR & $F_{\rm{var}}$ for the source in the NUV band \\
NUV\_VAR\_ERR & $1 \sigma$ error on NUV\_VAR \\
FUV\_RA\_J2000 &  Right ascension in degrees (in the FUV band)\\
FUV\_DEC\_J2000 &  Declination in degrees (in the FUV band)\\
A\_FUV & semi-major axis of the elliptical Kron aperture\footref{footnote_1} in arcsec (in FUV band)\footref{footnote_2}\\
B\_FUV & semi-minor axis of the elliptical Kron aperture\footref{footnote_1} in arcsec (in FUV band)\footref{footnote_2}\\
THETA\_FUV & Angle for the elliptical Kron aperture\footref{footnote_1} in degrees (in FUV band)\footref{footnote_2}\\
FUV\_FLUX   &   Estimated Flux of the source in the FUV band in units of  ergs ${\rm{cm}}^{-2} {\rm{sec}}^{-1}$ \AA${}^{-1}$\\
FUV\_FLUX\_ERR    &  Error on FUV\_FLUX  \\
FUV\_MAG    &  Estimated AB magnitide  of the source in the FUV band\\
FUV\_MAG\_ERR &   Error on FUV\_MAG \\
FUV\_SNR & Signal to Noise ratio of the source in the FUV band \\
FUV\_VAR & $F_{\rm{var}}$ for the source in the FUV band \\
FUV\_VAR\_ERR & $1 \sigma$ error on FUV\_VAR \\
UV\_COLOR & Difference between the FUV band magnitude and the NUV band magnitude\\
GAIA\_ID & ID of the source as per \gaia{} DR3 catalogue \\
GAIA\_RA\_J2016 &  Right ascension in degrees as per \gaia{} DR3 catalogue \\
GAIA\_DEC\_J2016 &  Declination in degrees as per \gaia{} DR3 catalogue  \\
GAIA\_FLUX   &   Flux of the source in \gaia{} DR3 catalogue in units of $e^{-} / s$\\
GAIA\_FLUX\_ERR    &  Error on GAIA\_FLUX  \\
GAIA\_MAG    &  AB magnitude  of the source calculated from GAIA\_FLUX \footnote{These magnitudes are calculated from GAIA\_FLUX applying the zero-point magnitude in AB magnitude system available at \url{https://www.cosmos.esa.int/web/gaia/edr3-passbands}}\\
GAIA\_CLASS &   Category of the source according to \gaia{} proper motion\footref{footnote_3} \\
GAIA\_PM & Total proper motion of the source in mas/year in \gaia{} DR3 catalogue \\
XMM\_ID & ID of the source in XMM DR12 catalogue \\
XMM\_RA\_J2000 &  Right ascension in degrees in XMM DR12 catalogue \\
XMM\_DEC\_J2000 &  Declination in degrees in XMM DR12 catalogue  \\
XMM\_FLUX   &   Flux of the source in XMM DR12 catalogue in units of  ergs ${\rm{cm}}^{-2} {\rm{sec}}^{-1}$\\
XMM\_FLUX\_ERR    &  Error on XMM\_FLUX  \\
REDSHIFT & Redshift of the source\footnote{`-9999' if not available \label{footnote_4}}\\
REDSHIFT\_ERR & Error on REDSHIFT\footref{footnote_4}\\
\hline
\end{tabular} 
\label{table:catalog_param}
\end{table*}

\subsection{Color-Color Diagram}\label{sect:color-color_diagram}
To identify candidates of different types of astronomical objects in our catalogue, we constructed color-color diagram using three different bands -- FUV, NUV and Optical (\gaia{}) magnitudes. 
 
\begin{figure}
\plotone{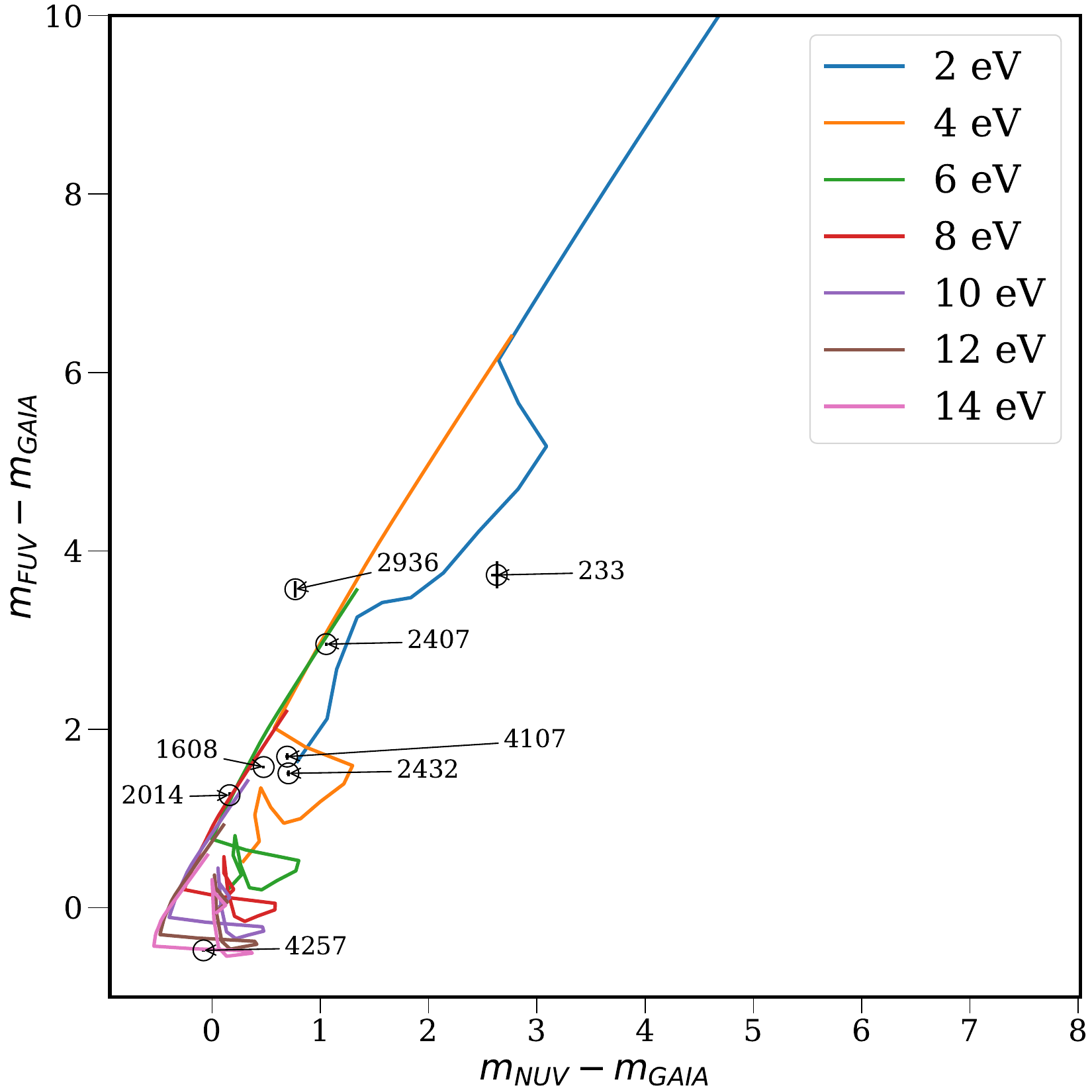}
\caption{Comparison of observed FUV-\gaia{} and NUV--\gaia{} colors with that predicted for type 1 AGN with a range of accretion disk temperatures $kT_{in}=2-14{\rm~eV}$ and redshifts $z=0.0-2.9$.
Similarity of the location of the observed colors and the simulated colors strongly suggests that the UVIT sources are type 1 AGN. The numbers annotated to the observed data points shows the ID for that source in our catalogue. \label{fig:color-color_plot}}

\end{figure}

We plot the FUV--GAIA and NUV--GAIA colors in Figure~\ref{fig:color-color_plot} for the point sources in our catalogue with \gaia{} counterparts that have no appreciable proper motion. In order to investigate if these sources are active galactic nuclei (AGN),
we simulated the $m_{FUV} - m_{\gaia{}}$ and $m_{NUV} - m_{\gaia{}}$ using typical spectral energy distribution of AGN. Since the emission from accretion disks in AGN completely dominates the optical/UV bands, we used accretion disk continuum for our simulations. We did not consider emission/absorption lines which are likely to affect the colors marginally. We used the multi-color accretion disk model {\tt diskbb} available in XSPEC (version 12.12.1) software \citep{1996ASPC..101...17A}.  We modified the {\tt diskbb} model to account for the Galactic reddening by multiplying with the {\tt redden} model component.  
 We obtained the color excess, $E(B-V)$,  from  the Galactic column density ($N_{H} = 4.10 \times 10^{20}  {\text{cm}}^{-2}$) along the line of sight to IC~4329A using the relation $N_{H} ({\text{cm}}^{-2}) = (6.86 \pm 0.27) \times 10^{21} E(B - V) (\text{mag})$ \citep{2009MNRAS.400.2050G}. We found  $E(B-V)= 0.059$ which we used in the {\tt redden} component. We then simulated several model spectra for the accretion disk  with the inner-disk temperature ($kT_{in}$) in the range of 2-14~eV and redshift in the range, $z=0.0-2.9$.  As we did not use intrinsic reddening, our simulated spectra represent only type 1 AGN. From the simulated accretion disk spectra modified by the Galactic reddening, we calculated expected count rates, FUV/NUV and \gaia{} G magnitudes in the AB system. 

\begin{figure}
\plotone{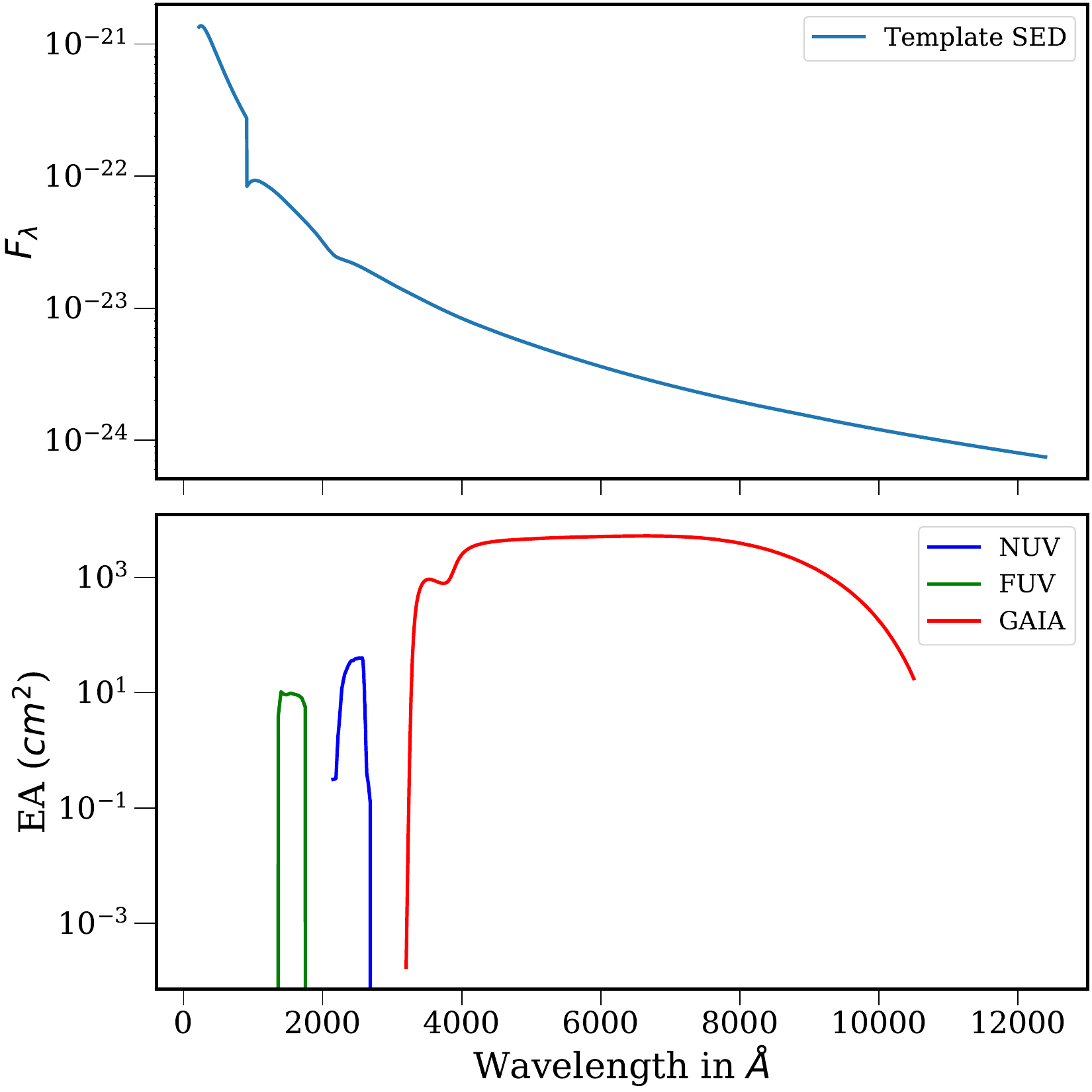}
\caption{{\it Upper panel:} Theoretical spectrum of an accretion disk derived from the XSPEC model {\tt diskbb} with an inner-disk temperature of 10~eV. $F_\lambda$ is flux density for an arbitrary normalisation of the {\tt diskbb} model. {\it Lower panel:} The effective area as a function of wavelength for the FUV/F154W, NUV/N245M and \gaia{} G band filter. The effective area for the \gaia{} G band is the product of the filter transmissivity and the telescope pupil area (see Section~4.2).\label{fig:color-color_explanation}}
\end{figure}

For the calculations of expected magnitudes in FUV and NUV bands, we  used the effective areas  for  the FUV filter (F154W) and NUV filter (N245M) and the magnitude zero-points given in \cite{2020AJ....159..158T}. We calculated the \gaia{} magnitude following the prescription described in \cite{2018gdr2.reptE....V}. First we calculated the expected flux in the G band  from the simulated disk spectrum by multiplying the spectrum with the filter transmissiivity\footnote{available at \url{https://www.cosmos.esa.int/web/gaia/edr3-passbands}} and telescope pupil area ($7278{\rm~cm^{2}}$) and integrating over the bandpass. The product of the filter transmissivity and the telescope pupil area which we refer as the effective area is plotted in the bottom panel of Figure~\ref{fig:color-color_explanation}. We converted the estimated \gaia{} flux to AB magnitude in the G band using the zero point magnitude of $25.8010$.  
Figure~\ref{fig:color-color_explanation} shows the model spectrum of an accretion disk with inner disk temperature of 10~eV, and the effective areas of the three filter band passes used here. To find out the reliability of our \gaia{} magnitude calculator, we used the UV/optical spectrum of WD~0308--565 which is a spectrophotometric standard\footnote{\url{https://www.stsci.edu/hst/instrumentation/reference-data-for-calibration-and-tools/astronomical-catalogs/calspec}}, and calculated its \gaia{} G magnitude to be $14.22$. The \gaia{} DR3 lists the G-band source flux for WD~0308--565 as 43019.27 $e^{-} s^{-1}$ corresponding to an AB magnitude of $14.21$ which is similar to our predicted magnitude from the spectrum. This establishes the reliability of our magnitude simulator.

For each inner-disk temperature and redshift, we plotted the difference of NUV and \gaia{} magnitudes i.e., NUV--\gaia{} color vs the difference of FUV and \gaia{} magnitude i.e., FUV--\gaia{} color along the y axis in Figure~\ref{fig:color-color_plot}. Thus the color-color plot we generated shows the expected color-color curves for accretion disk emission for a range of inner-disk temperatures and redshifts. The circular data points in Figure~\ref{fig:color-color_plot} show the measured values of NUV - \gaia{} color and FUV - \gaia{} color of 
the point sources in our catalogue that have \gaia{} counterparts but no appreciable proper motion i.e, they are unlikely to be stars. The location of the observed color-color data points within or near the color-color courves derived for the hypothetical AGN clearly demonstrate that these sources are most likely type 1 AGN. We have mentioned details of these sources in Table \ref{Table:cand_AGN_table}. We could identify two of these 8 sources in Million Quasars Catalogue, version 8 \citep{2023OJAp....6E..49F}. These two sources are source ID: 1608 and source ID: 2407, identified as the Quasar 1WGA J1349.6-3016 and Q 1345--301 at redshift 0.543 and 1.438 respectively in Million Quasars catalogue.

\begin{deluxetable*}{ccccccc}
\tablecaption{Details of the identified candidate AGNs in our catalogue
\label{Table:cand_AGN_table}}
\tablehead{\colhead{ID} & \colhead{RA (J2000)} & \colhead{DEC (J2000)} & \colhead{NUV flux}  & \colhead{FUV flux}   & \colhead{X-ray flux}  & \colhead{Redshift}\\
&(Degrees) &(Degrees) & ($10^{-17}$ ergs ${\rm{cm}}^{-2} {\rm{s}}^{-1}$ \AA${}^{-1}$) & ($10^{-17}$ ergs ${\rm{cm}}^{-2} {\rm{s}}^{-1}$ \AA${}^{-1}$) & ($10^{-13}$ ergs ${\rm{cm}}^{-2} {\rm{s}}^{-1}$)}
\startdata
233 & 207.208859 & -30.145217 & 0.70 $\pm$ 0.03 & 0.65 $\pm$ 0.08  & -- & --\\
2407 & 207.132445 & -30.356413 & 34.4 $\pm$ 0.2 & 14.6 $\pm$ 0.3  & 0.7 $\pm$ 0.2 & 1.438\\
2936 & 207.27747 & -30.385212 & 6.76 $\pm$ 0.08 & 1.3 $\pm$ 0.1  & 10.4 $\pm$ 0.2 & --\\
4107 & 207.307791 & -30.318886 & 4.68 $\pm$ 0.07 & 4.7 $\pm$ 0.2  & -- & --\\
2014 & 207.465311 & -30.311691 & 6.40 $\pm$ 0.08 & 6.2 $\pm$ 0.2  & 0.4 $\pm$ 0.1 & --\\
1608 & 207.414134 & -30.273163 & 23.0 $\pm$ 0.2 & 22.8 $\pm$ 0.3  & 2.5 $\pm$ 0.3 & 0.543\\
2432 & 207.292723 & -30.345685 & 4.92 $\pm$ 0.07 & 6.0 $\pm$ 0.2  & 0.20 $\pm$ 0.03 & --\\
4257 & 207.180977 & -30.207083 & 11.8 $\pm$ 0.1 & 36.7 $\pm$ 0.4  & -- & --\\
\enddata
\end{deluxetable*}

\subsection{Variability of sources}
We also measured the FUV and NUV flux of our detected sources in each of the five UVIT observations. We used the same source extraction regions we had obtained earlier for the merged image using the {\tt SExtractor} tool, and performed photometry of all sources.   
We generated the light curves of the point sources using the count rates measured with the five observations. In order to investigate UV variability of point sources,  we calculated the fractional excess variability amplitude ($F_{\rm{var}}$) of the light curve of each source in our catalog. This quantity gives an estimate of the intrinsic variability amplitude of source relative to the mean count rate  after removal of the contribution of the measurement errors, and thus gives an estimate of the intrinsic variability amplitude of the sources \citep{2002ApJ...568..610E}. The $F_{\rm{var}}$ is expressed as \citep{2002ApJ...568..610E}
\begin{equation}
\centering
F_{\rm{var}}=\sqrt{{S^{2}-{\bar{\sigma^{2}}_{\rm{err}}}\over\bar{x}^2}}
\end{equation}
where $S^2$ represents the variance in the binned data comprising the light curve, $\bar{x}$ is the mean count rate i.e. arithmetic mean of $x_i$ (the count rate measurements for each observation) 
and $\bar{\sigma^{2}}_{\rm{err}}$ represents the mean square error in count rate measurement. 
$S^2$ and ${\bar{\sigma^{2}}_{\rm{err}}}$ are given as
\begin{equation}
\centering
S^2={1\over{N-1}}\sum_{i=1}^{N}(x_i-\bar{x})^2
\end{equation}
\begin{equation}
\bar{\sigma^{2}}_{\rm{err}}={1\over{N}}\sum_{i=1}^{N}\sigma^{2}_{\rm{err,i}}
\end{equation}
 
The $1\sigma$ uncertainty in the fractional variability amplitude is given by

\begin{equation}
  \centering
           err(F_{\rm{var}})=\frac{1}{F_{var}} \sqrt{\frac{1}{2N}} \frac{S^2}{{\bar{x}}^2}
 \end{equation}

Here we consider a source as variable if its  $F_{\rm{var}}$ is more than $2.5 \times err(F_{\rm{var}})$. We list all such sources in Table~\ref{Table:vartable}.

We found a total of 28 sources that are variable in the NUV band. There are only three FUV sources with $F_{\rm{var}} > 2.5 \times err(F_{\rm{var}})$ (see Table~\ref{Table:vartable}).  We quote the $2\sigma$ upper-limit on the $F_{\rm{var}}$ in the FUV band if a source is not variable as  per our criterion. We also provide information on \gaia{} and X-ray counterparts. 
16 of these NUV variable sources are \gaia{} stars. We also checked if  background was variable in the five observations. For this purpose we used 20 circular regions of $10\arcsec$ radius placed in the source-free regions across the field and calculated $F_{\rm {var}}$ for each of 20  regions. For all these 20 regions we found that our background was not variable. We identified 6 out of these 28 variable sources as AGN candidates in section \ref{sect:color-color_diagram}. We mentioned these sources as ``AGN Candidate (UVIT)" in column 6 of table \ref{Table:vartable}. We could also identify 8 sources in SIMBAD catalogue \citep{2000A&AS..143....9W}. We added the SIMBAD classification of these sources in the column 6 of Table \ref{Table:vartable}. We could also identify 22 out of these 28 sources in \gaia{} DR3 catalogue and 3 of them were classified as QSO candidates. We have added the \gaia{} classification as well in the column 6 of table \ref{Table:vartable}. 
Based on the effective temperature measured by Gaia for 13 stars in our sample, we made tentative identification as follows: one of them is an F-type star (id: 2902), two are K-type stars (ids: 2710, 3265) and the rest of them are G-type stars (ids: 67, 683, 1916, 2470, 2667, 3151, 3195, 3549, 3566, 4307). We found that the 17 sources, that are identified to be most likely stars, have fractional variability amplitude (in the NUV band) in the range of $1.3-67.8\%$ and the 8 sources, identified as likely to be AGNs or QSOs, have fractional variability amplitude (in NUV) in the range of $5.8-17.2\%$.
 
Although  the significance for variability in the range of $2.5-3\sigma$ is not great and using only five observations we can not comment on the periodicity of the variable sources, the excess variability amplitude values we found are well within the range found by \cite{2024A&A...687A.313B} in a variable UV sky survey using {\it GALEX} data on 4202 variable sources. Despite a low number of sources in our sample, we could see higher variability amplitude for stellar objects than AGNs which is a similar trend found by \cite{2024A&A...687A.313B} and \cite{2013ApJ...766...60G}. We expect the variable stars in our catalogue to be mostly of the RR Lyrae type, which are predominant in UV-variable stellar classes, as found in previous studies \citep{2024A&A...687A.313B, 2013ApJ...766...60G}. It is also possible that some of the variable stars are of other type such as Cephieds or binary systems, possibly with white dwarfs. Detailed spectroscopic studies  will confirm the stellar types.

\begin{figure*}
\plotone{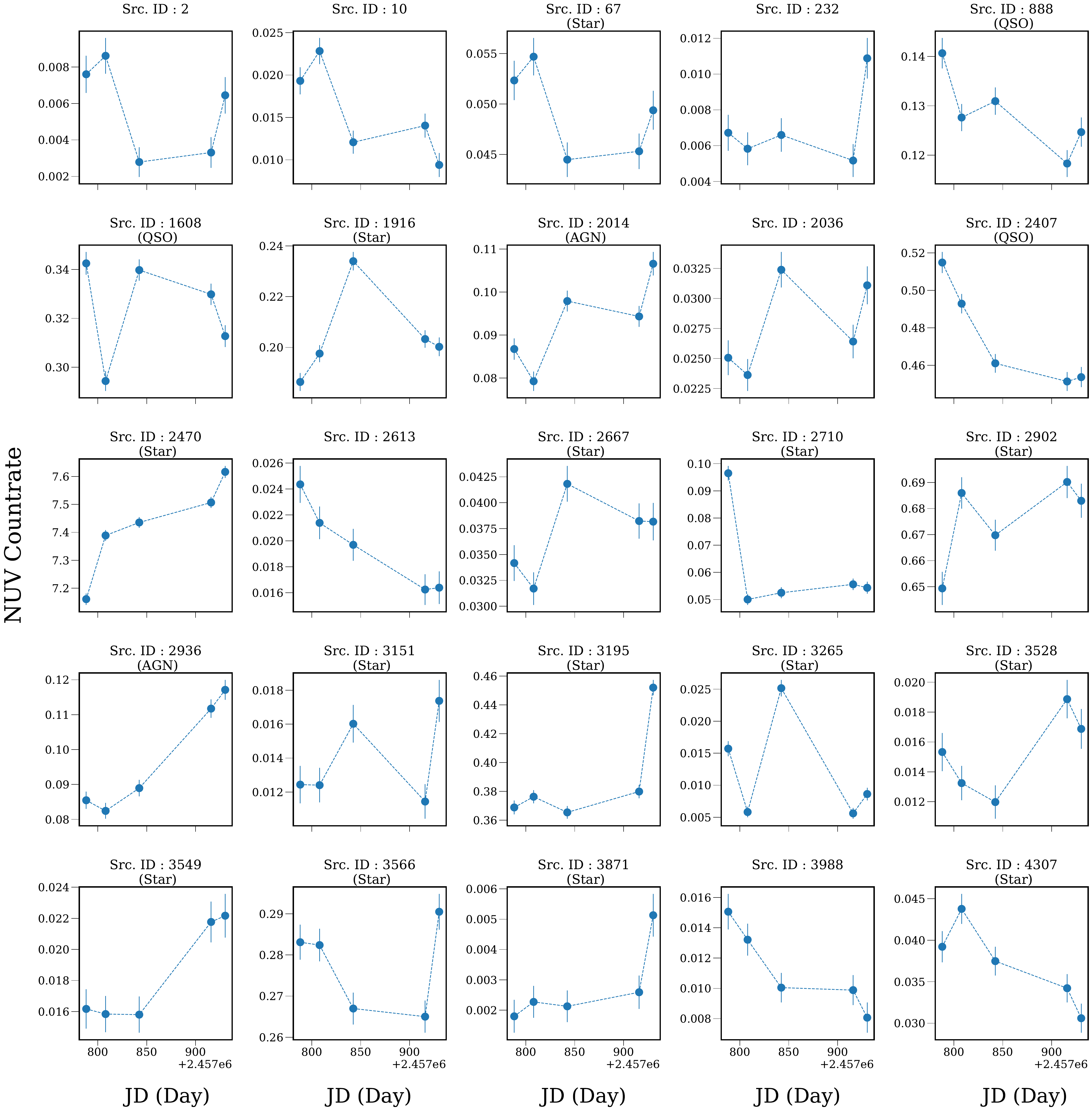}
\caption{NUV light curves of 25 point sources that are variable in NUV above $2.5 \sigma$ significance level but are not variable in FUV band. Julian date of our observation is shown on the x axis and the NUV count rate for the sources are plotted along y axis. In the header of each plot the source ID is mentioned along with their probable classification in the bracket. \label{fig:variance}}
\end{figure*}

\begin{figure*}
\plotone{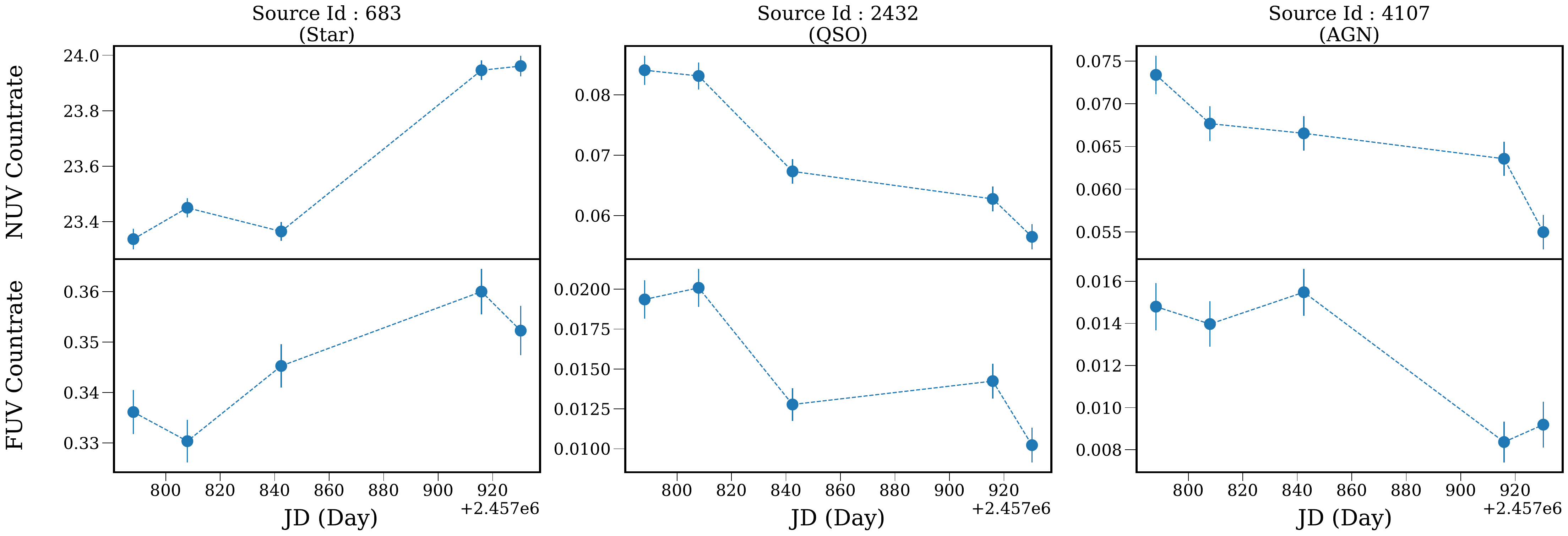}
\caption{light curves of 3 point sources that are variable above $2.5 \sigma$ significance level in both the NUV and the FUV bands. Upper Panel: NUV light curves for these three sources. Lower panel : FUV light curve of the corresponding source. In the header of each panel the source ID is mentioned along with their probable classification in the bracket.\label{fig:fuv_nuv_variance}}
\end{figure*}

\begin{table*}
\caption{Variability of point-like sources that are variable above $2.5 \sigma$ level in NUV band.}
\begin{center}
\begin{tabular}{m{1cm} m{1cm} m{1cm} m{1cm} m{1cm} m{4.5cm} m{2.5cm}}
\hline \\
Source ID & $\left(F_{\rm{var}}\right)$ in NUV &  Error on $\left(F_{\rm{var}}
\right)$ in NUV & $\left(F_{\rm{var}}\right)$ in FUV &  Error on $\left(F_{\rm{var}}
\right)$ in FUV & Classification & X-ray flux (0.2-12 keV) \\
 & & & & & & ($10^{-14} \rm{ergs}~ {cm}^{-2} s{-1}$)\\ 
\hline
2 & 0.420 & 0.153 & --- & ---  & --- & ---  \\ 
10 & 0.339 & 0.115 & --- & ---  & --- & ---  \\ 
67 & 0.081 & 0.031 & --- & ---  & star (\gaia{})  & ---  \\ 
232 & 0.285 & 0.112 & --- & ---  & --- & ---  \\ 
683 & 0.013 & 0.004 & 0.032 & 0.012  & star (SIMBAD, \gaia{}) & ---  \\ 
888 & 0.060 & 0.022 & --- & ---  & star (\gaia{}) & ---  \\ 
1608 & 0.061 & 0.020 & --- & ---  & QSO (SIMBAD, \gaia{}), AGN candidate (UVIT) & 25.11  \\ 
1916 & 0.085 & 0.028 & --- & ---  & star(\gaia{}) & ---  \\ 
2014 & 0.110 & 0.037 & --- & ---  & AGN candidate (SIMBAD, UVIT) & 4.16  \\ 
2036 & 0.128 & 0.047 & --- & ---  & --- & 1.48  \\ 
2407 & 0.058 & 0.019 & --- & ---  & QSO (SIMBAD, \gaia{}), AGN candidate (UVIT) & 6.79 \\ 
2432 & 0.172 & 0.056 & 0.268 & 0.091 & X-ray source (SIMBAD), AGN candidate (UVIT) & 2.04 \\ 
2470 & 0.023 & 0.007 & --- & ---  & star(\gaia{}) & 13.57 \\ 
2613 & 0.163 & 0.060 & --- & ---  & ---  & ---  \\ 
2667 & 0.096 & 0.038 & --- & ---  & star(\gaia{})  & ---  \\ 
2710 & 0.314 & 0.100 & --- & ---  & star(SIMBAD ,\gaia{}) & 7.47\\ 
2902 & 0.023 & 0.008 & --- & ---  & star(\gaia{}) & ---  \\ 
2936 & 0.163 & 0.053 & $<$0.515 & --- & X-ray source (SIMBAD), QSO(\gaia{}), AGN candidate (UVIT) & 10.45  \\ 
3151 & 0.169 & 0.065 & --- & --- & star(\gaia{}) & --- \\ 
3195 & 0.092 & 0.030 & --- & --- & star(\gaia{}) & --- \\ 
3265 & 0.678 & 0.218 & --- & --- & star(\gaia{}) & 0.66 \\ 
3528 & 0.162 & 0.064 & --- & --- & star(\gaia{}) & --- \\ 
3549 & 0.167 & 0.062 & --- & --- & star(\gaia{}) & --- \\ 
3566 & 0.037 & 0.014 & --- & --- & star(\gaia{}) & 1.64 \\ 
3871 & 0.437 & 0.169 & --- & --- & star(\gaia{}) & ---  \\ 
3988 & 0.233 & 0.085 & --- & --- & --- & ---  \\ 
4107 & 0.098 & 0.034 & 0.255 & 0.090 & star(\gaia{}), AGN candidate (UVIT) & ---  \\ 
4307 & 0.126 & 0.046 & --- & ---  & star(\gaia{}) & ---  \\
\hline
\end{tabular} 
\label{Table:vartable}
\end{center}
\end{table*}

\subsection{Candidate white dwarfs in the field}
To find the white dwarfs in our field we cross-matched our catalogue with the white dwarf catalogue produced by \citep{2021MNRAS.508.3877G} based on \gaia{} EDR3 (Early Data Release 3) database.  \citet{2021MNRAS.508.3877G} presented the largest white dwarf catalogue containing 1280266 white dwarf candidates and  also assigned a probability for each candidate  to be a white dwarf ($P_{WD}$ column in their catalogue) and derived different physical parameters for the high-confidence white dwarf candidates. By cross-matching this catalogue  with our catalogue we found two sources with \gaia{} source id 6175211520428699776 and 6175157296466199168. We refer them  as white dwarf candidate 1 and 2, respectively. The probabilities of them being a white dwarf are 0.265756 and 0.149864 (as mentioned in the white dwarf catalogue), respectively. Since white dwarfs emits mostly in UV, we used the additional UV band data available to us for these two sources to confirm whether they really are white dwarfs. We used the measured fluxes and magnitudes of these two sources in the four bands: our NUV band (2167-2727 \AA) and three \gaia{} bands, namely \gaia{} G band (3200-10500 \AA), $G_{BP}$ band (3250-7500 \AA) and $G_{RP}$ band (6100-10800 \AA). We first derived four colors $m_{G} - m_{G_{BP}}$, $m_{G}- m_{G_{RP}}$, $m_{G_{BP}}- m_{G_{RP}}$ and $m_{G}-m_{NUV}$ using the AB magnitudes in the four bands. Next we compared these observed colors with their theoretical values using Koester white dwarf model spectra\footnote{available at \url{http://svo2.cab.inta-csic.es/theory/newov2/index.php?models=koester2}}, described in \citep{2010MmSAI..81..921K} and \citep{2009ApJ...696.1755T}. These models provide spectra of white dwarfs with Hydrogen-rich atmosphere for a wide range of effective temperature (5000--80000~K) and  surface gravity ($\log{g}$ in the range of 6.5--9.5). 
We obtained a total of 1066 spectra and folded them with the filter response of the four bands to estimate flux and colors.
We compared the model predicted colors with the observed colors of each of the two candidate stars in our catalogue and found the best-fit model for  the lowest chi-square. The best-fit white dwarf model for the two stars are $T_{eff} = 6250{\rm~K}$  and $\log{g} = 9.0$, and $T_{eff} = 6500{\rm~K}$ and $\log{g} = 8.75$ for the white dwarf candidate 1 and 2, respectively. By normalising these best-fit model  spectra to match our measured flux we found the ratio R/D where R is the stellar radius for the white dwarf and D is the distance. Using the distance calculated from parallax measured by \gaia{} we derived the radii  to be 29232.2~km and 45636.8~km for the candidates 1 and 2, respectively. Thus using the additional photometric data from our NUV band, we conclude that none of these two stars are white dwarfs.

\subsection{Sources with unusual morphology}\label{sect:int_sources}
\begin{figure*}
\gridline{\fig{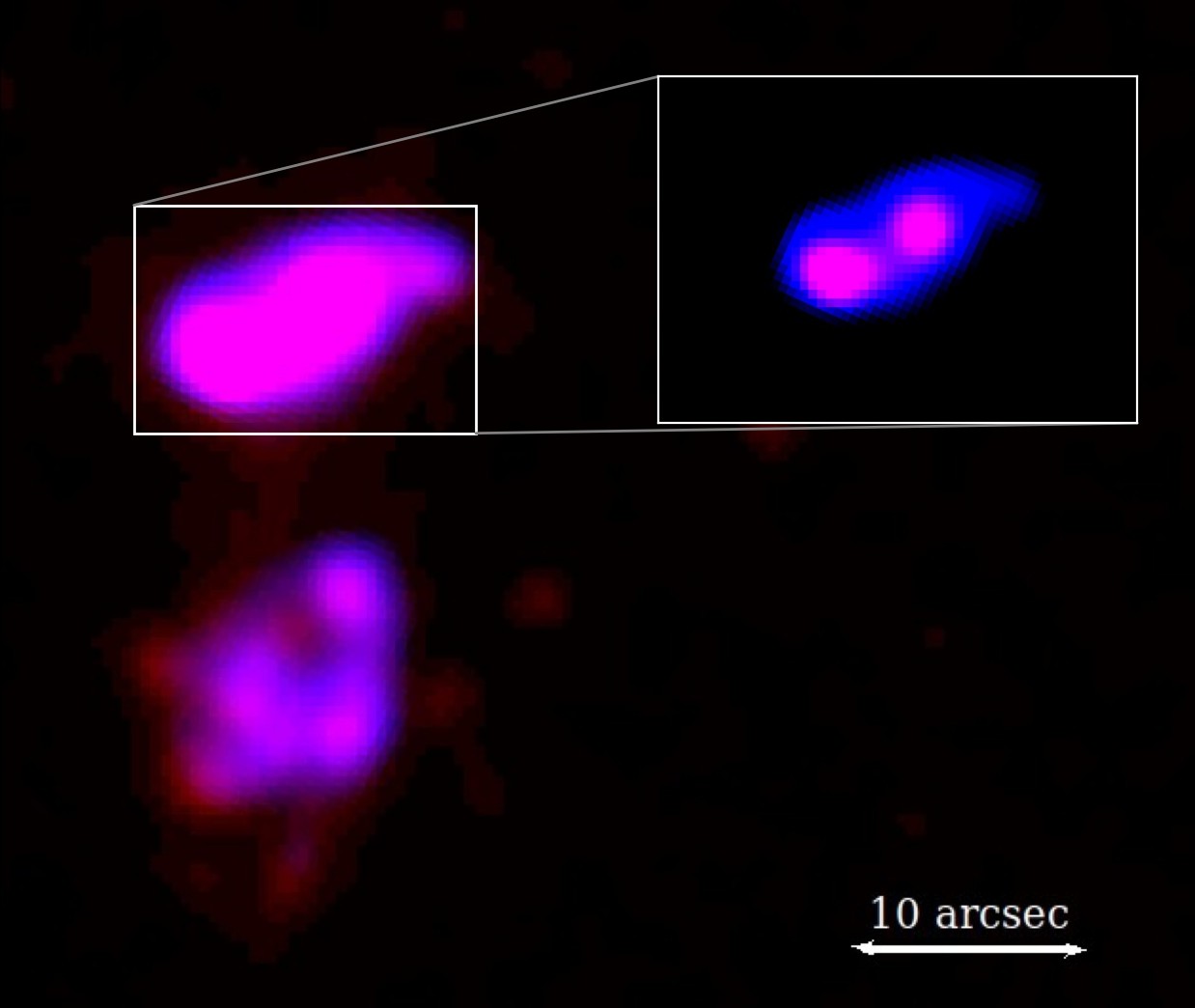}{0.42\textwidth}{(a) LEDA 719417 (Source ID : 3915) and \\LEDA 719364 (source ID : 3909)}
           \fig{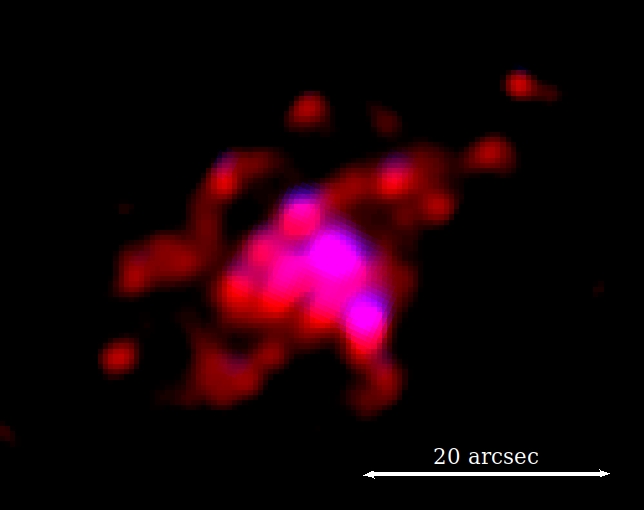}{0.45\textwidth}{(b)IC 4327 (Source ID : 4255)} 
            }
\gridline{\fig{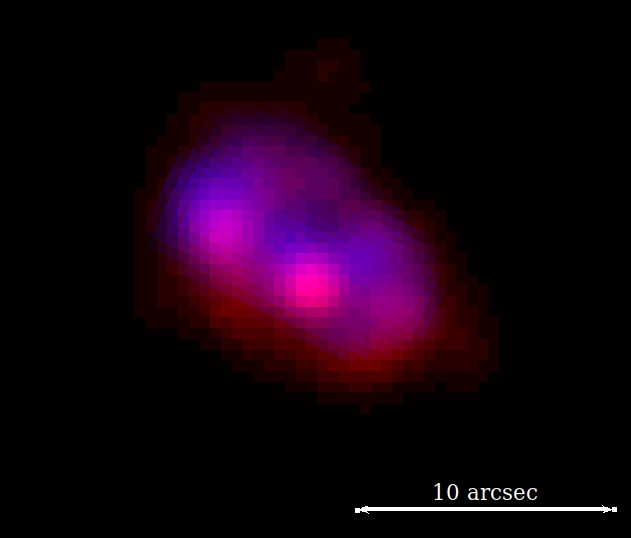}{0.45\textwidth}{(c) LEDA 720905 (Source ID : 4182)}
            \fig{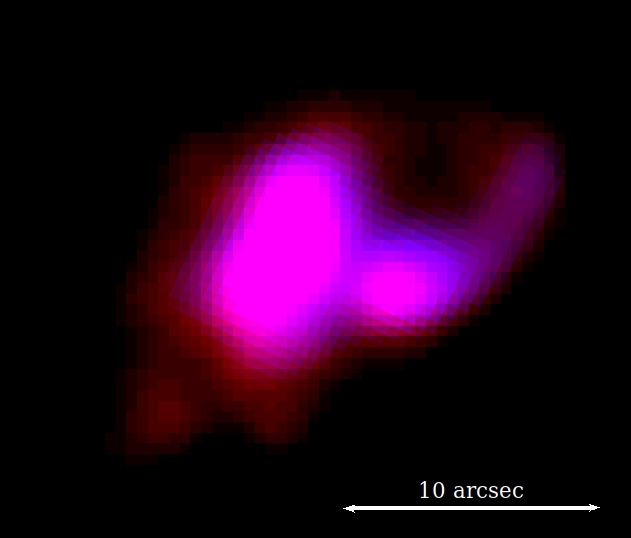}{0.45\textwidth}{(d) FLASH J134835.63-302441.5 (Source ID : 4165)}
            }
\gridline{\fig{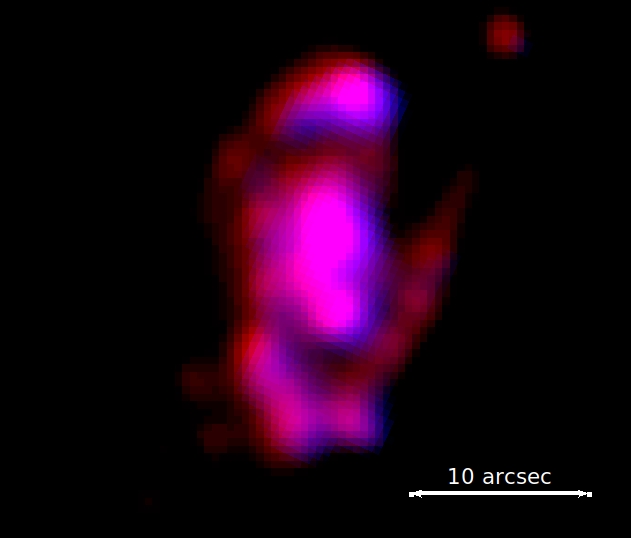}{0.45\textwidth}{(e) LEDA 89764 (Source ID : 4173)}
            \fig{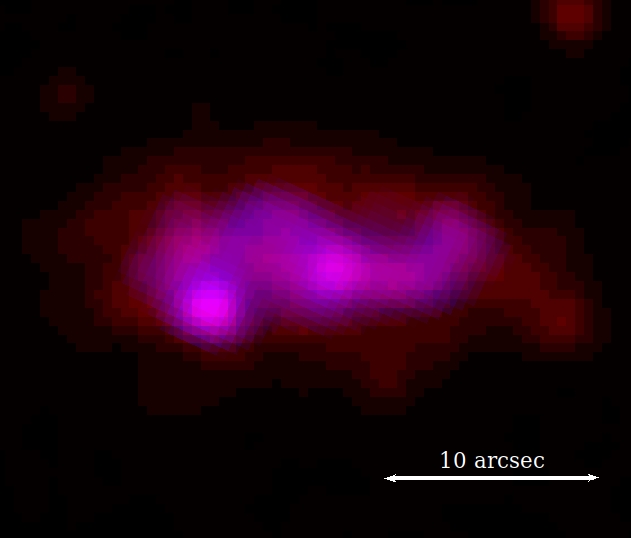}{0.45\textwidth}{(f) WISEA J134954.43-301349.0 (Source ID : 4180)}
            }
\caption{two colour images of few sources in our catalogue with interesting morphological features. In all the images NUV waveband image (smoothed using a Gaussian filter with FWHM of $1.2^{\prime \prime}$) is shown in red colour and FUV band image (smoothed using a Gaussian filter with FWHM of $2.0^{\prime \prime}$) is shown in blue colour. \label{fig:int_srcs}}
\end{figure*}

There are a few sources in our field that exhibit unusual morphology. We show the images of these sources in Figure~\ref{fig:int_srcs} and discuss their possible nature by inspecting their morphology.

\begin{enumerate}
    \item Interacting/merging galaxies: As shown in Figure~\ref{fig:int_srcs}(a), we detected two extended sources separated only by $\sim9\arcsec$, these sources (ID: 3915 at $\alpha = 207.319856^{\circ}$, $\delta = -30.264792^{\circ}$ and ID: 3909 at $\alpha = 207.319865^{\circ}$, $\delta=-30.260438^{\circ}$) are likely in a group and undergoing interactions.  The source 3909 seems to be a spiral galaxy with total NUV flux $19.1\pm0.2 \times 10^{-17}$ ergs ${\rm{cm}}^{-2} {\rm{s}}^{-1}$ \AA${}^{-1}$ and magnitude of 20.0 and FUV flux of $30.5\pm 0.5\times 10^{-17}$ ergs ${\rm{cm}}^{-2} {\rm{s}}^{-1}$ \AA${}^{-1}$ and magnitude 20.3.
    This source is most likely undergoing intense star formation causing multiple star forming regions as seen in the image. We identify this source with the galaxy LEDA~719417 which is seen as a diffuse source in the DSS images.
    
   The source 3915 ($\alpha = 207.319856^{\circ}$, $\delta = -30.264792^{\circ}$) is detected as an extended source in both NUV and FUV bands  with fluxes in respective bands as $35.2\pm 0.2\times 10^{-17}$ ergs ${\rm{cm}}^{-2} {\rm{s}}^{-1}$ \AA${}^{-1}$ and $54.8\pm0.5\times 10^{-17}$ ergs ${\rm{cm}}^{-2} {\rm{s}}^{-1}$ \AA${}^{-1}$. The AB magnitudes of the source in NUV and FUV band respectively are 19.3 1nd 19.7. This source shows two bright compact sources, separated by $\sim 5\arcsec$ only,  which are clearly seen in the inset in Figure~\ref{fig:int_srcs}(a).  It also shows hints of spiral arms, hence the source is most likely  a merging system. The two bright compact sources appear to be the nuclei (probably active) of the merging galaxies, hence the source can be classified as a galaxy with dual nuclei. The source seen on the left of the inset image has an NUV and FUV flux of $14.2\pm 0.1\times 10^{-17}$ ergs ${\rm{cm}}^{-2} {\rm{s}}^{-1}$ \AA${}^{-1}$ and $21.9\pm 0.3\times 10^{-17}$ ergs ${\rm{cm}}^{-2} {\rm{s}}^{-1}$ \AA${}^{-1}$ while the source seen on the right has a flux of $11.3\pm 0.1\times 10^{-17}$ ergs ${\rm{cm}}^{-2} {\rm{s}}^{-1}$ \AA${}^{-1}$ in NUV and a flux of $17.0\pm 0.3\times 10^{-17}$ ergs ${\rm{cm}}^{-2} {\rm{s}}^{-1}$ \AA${}^{-1}$ in FUV. 
  We could identify a radio counterpart in NVSS catalogue that might be associated with this interacting system. We have mentioned the details of this radio counterpart in Table \ref{table:radio_counterpart}. This radio counterpart has a spectral flux density of $2.9 \pm0.6$ mJy at 1.4 GHz.
   Optical spectroscopy of the nuclei is required to reveal the nature of the likely merging system. We identify this dual nuclei galaxy with LEDA~719364 which appears as an elongated structure in the DSS images. Hence, our UVIT observations discovered the dual nuclei in this galaxy.

    \item IC~4327: The source (ID: 4255) shown in the figure~\ref{fig:int_srcs}(b) at the sky position ($\alpha=207.182512^{\circ}$, $\delta=-30.217489^{\circ}$) is identified as the galaxy IC~4327. This is classified as an optical emission line galaxy. The galaxy is detected in the NUV band with a magnitude of 16.2 and total NUV Flux of $593.3\pm0.8\times 10^{-17}$ ergs ${\rm{cm}}^{-2} {\rm{s}}^{-1}$ \AA${}^{-1}$ as well as in the  FUV band with a magnitude of 16.6 and FUV flux of $93.5\pm0.2\times 10^{-16}$ ergs ${\rm{cm}}^{-2} {\rm{s}}^{-1}$ \AA${}^{-1}$. The UV band profile of the galaxy shows three distinct blob like structure that emits more strongly than other regions of the galaxy both in NUV and FUV band. The UV-rich region in the central part of the galaxy is most likely the result of intense star formation in the nuclear regions, there may be contamination due to presence of an active nucleus. 
    Given the fact that arms of  spiral galaxies generally inhabit star forming regions that emit primarily in the UV band, the two distinct blob like strong UV emitting regions in the spiral arms of the galaxy might be arising due to star forming regions. We also identified the radio counterpart of this galaxy in NVSS catalogue (see Secton \ref{sect:multiband_counterparts} and Table \ref{table:radio_counterpart}) with a spectral flux density of $2.9 \pm 0.6 $ mJy at 1.4 GHz. 
    Further study with multi-band and spectroscopic data may reveal details on the UV emitting regions in the galaxy.
    
    \item LEDA~720905: We discover a ring-like structure with a size $\sim 4-5\arcsec$ around the bright nucleus in the source (ID: 4182, sky position: $\alpha=207.36582^{\circ}$, $\delta=-30.141252^{\circ}$) shown in the Figure~\ref{fig:int_srcs}(c) which we identify with the galaxy LEDA~720905 located at $z=0.076654\pm0.000130$. We measured NUV flux of $17.3 \pm  0.1\times 10^{-17}$ ergs ${\rm{cm}}^{-2} {\rm{s}}^{-1}$ \AA${}^{-1}$ and FUV flux of $20.7\pm0.4\times 10^{-17}$ ergs ${\rm{cm}}^{-2} {\rm{s}}^{-1}$ \AA${}^{-1}$ . The source is detected with NUV magnitude of 20.1 and FUV magnitude of 20.9. The morphological features shown in the two-colour image of the source suggests the possibility of an active galactic core emitting strongly in both Far and Near UV. The ring-like structure around the nucleus could represent the spiral arms of the galaxy. 
    
    \item FLASH~J134835.63-302441.5: The source  (ID: 4165) shown in the Figure~\ref{fig:int_srcs}(d) is located at the sky position ($\alpha=207.148182^{\circ}$, $\delta=-30.4116^{\circ}$)  with NUV magnitude of 18.5 and FUV magnitude of 19.0. We identify this source as the galaxy FLASH~J134835.63-302441.5 at $z=0.013534\pm0.000110$. The source has a total NUV flux of $71.9\pm0.3\times 10^{-17}$ ergs ${\rm{cm}}^{-2} {\rm{s}}^{-1}$ \AA${}^{-1}$ and total FUV flux of $118.0\pm 0.8\times 10^{-17}$ ergs ${\rm{cm}}^{-2} {\rm{s}}^{-1}$ \AA${}^{-1}$. The source shows two extended structures very close to each other.  These complex structures are detected in both the FUV and NUV bands and may be the result of two merging/merged spiral galaxies.
    
    \item LEDA~89764: The source shown in Figure~\ref{fig:int_srcs}(e) is located at a sky position $\alpha=207.274258^{\circ}$, $\delta=-30.479603^{\circ}$). It is inside the field of coverage common to all five observations in the NUV band but not all five observations in FUV band have covered this source. This is why we mention only the NUV properties of the source (ID~4173). The source is detected as an extended NUV source with flux of $135.1\pm0.4\times 10^{-17}$ ergs ${\rm{cm}}^{-2} {\rm{s}}^{-1}$ \AA${}^{-1}$ and magnitude of 17.8 . The morphological features shown in the two-colour image of the source consist of multiple UV emitting regions -- the central  region which itself is extended with complex structure and the two outer regions possibly the spiral arms. We identify this source with the optical emission line galaxy LEDA~89764.
    
    \item WISEA J134954.43-301349.0: The extended source shown in Figure~\ref{fig:int_srcs}(f) (source ID : 4180) is located at a sky position $\alpha=207.477842^{\circ}$, $\delta=-30.230305^{\circ}$ with NUV flux of $23.3\pm0.2\times 10^{-17}$ ergs ${\rm{cm}}^{-2} {\rm{s}}^{-1}$ \AA${}^{-1}$ and FUV flux of $32.3\pm 0.5\times 10^{-17}$ ergs ${\rm{cm}}^{-2} {\rm{s}}^{-1}$ \AA${}^{-1}$. The source is detected with magnitude of 19.7 in NUV band and magnittude of 20.4 in FUV band. The source appears to be a galaxy possibly with spiral arms or complex structures due to merger/interaction. We indentify this galaxy with WISEA J134954.43-301349.0. 

    \end{enumerate}

\subsection{Some interesting sources outside our field of coverage}\label{sect:int_sources_2}
We have found two interesting galaxies outside our field of coverage. Since the patch of the sky where they are located was not covered by all five observations mentioned in section \ref{sect:data_reduction}, we have not included them in our catalog, but still we would like to bring out their interesting morphological features in the FUV and NUV bands.
\begin{figure*}
\gridline{\fig{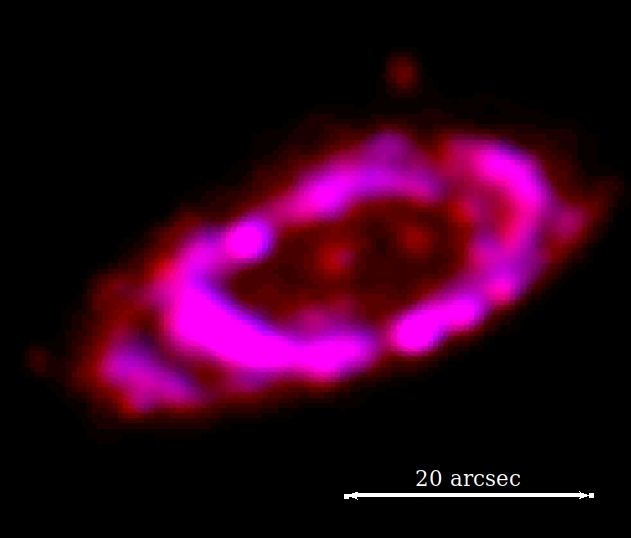}{0.45\textwidth}{}
        \fig{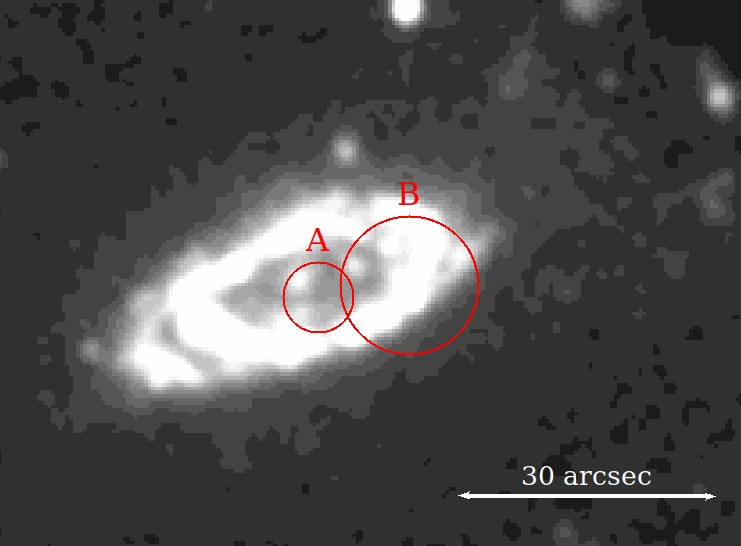}{0.52\textwidth}{}
}
\caption{\astrosat{}/UVIT images of NGC~5298. {\it Left panel} : Two-color image constructed from  NUV band (red) and FUV band (blue)
images. A ring-like structure possibly connected with spiral arms and point-like sources inside the ring are clearly seen. {\it Right panel} : NUV band image of the same source showing extended diffuse emission. The two \xmm{} sources found inside the galaxy are shown in red-coloured circles with labels `A' and `B'. The radius of each circle correspond to 3 $\sigma$  where $\sigma$ represents the astrometric error on the X-ray source position. Additional diffuse emission including a patchy region in the upper-right side of the galaxy is clearly seen. \label{fig:int_srcs_out_1}}
\end{figure*} 
    
\begin{enumerate}
    \item NGC~5298: A galaxy with a ring-like structure, located at a at sky position $\alpha$(J2000)$=207.1518487^{\circ}$, $\delta$(J2000)$=-30.4282587^{\circ}$, is shown in the Figure~\ref{fig:int_srcs_out_1}. We identify this source with NGC~5298 at $z=0.01452$ using the  SIMBAD astronomical database \citep{2000A&AS..143....9W}. This galaxy is known to host a Seyfert 2 type AGN. 
    The galaxy shows strong UV emitting elliptical ring-like structure with devoid of UV emission in the inner regions except for two point-like source. The central point-like source is most likely the Seyfert 2 AGN. We measured the UV flux of the central source to be $2.9 \pm 0.1 \times 10^{-17}{\rm~ergs~cm^{-2}~s^{-1}}$\AA$^{-1}$ in the NUV band and $3.3\pm0.4 \times 10^{-18}{\rm~ergs~cm^{-2}~s^{-1}}$ \AA$^{-1}$ in the FUV band. By cross matching with the \xmm{}  source catalogue, we found two X-ray sources within this galaxy, one near the central point-like source i.e., the type 2 Seyfert, and the other near the elliptical ring. In Figure~\ref{fig:int_srcs_out_1} (right), we show the NUV image of the galaxy, we mark  the two X-ray sources detected with \xmm{} with red circles. The radii of the red circles correspond to the 3 $\sigma$ sizes where $\sigma$ is astrometric error on the  X-ray source position obtained from the \xmm{} source catalogue.  The \xmm{} sources are labelled as `A' and `B' in the figure and their \xmm{} IDs are 201474401010035 and 208007608010045, respectively, with X-ray (0.2-12 KeV) fluxes of $(1.2 \pm 0.5) \times 10^{-14}{\rm~ergs~{cm^{-2}} {s}^{-1}}$ and $(1.9 \pm 1.0) \times 10^{-14}{\rm~ergs~{cm^{-2}} {s}^{-1}}$. Using the redshift of the galaxy, we found that the central source i.e., the type 2 Seyfert has an X-ray (0.2-12 keV) luminosity of  $(5.0 \pm 2.1)\times10^{39}{\rm~ergs~ {s}^{-1}}$. The source labelled as `B' in the figure is found to have an X-ray (0.2-12 keV) luminosity of $(7.5 \pm 4.2)\times10^{39}{\rm~ergs~ {s}^{-1}}$. 

    In case of type 2 Compton-thick AGN, the primary X-ray emission can be completely absorbed below $\sim10$~keV, and the observed X-ray flux can be entirely the scattered emission which is generally  a few percent of the  intrinsic X-ray flux. In case of NGC~5298, assuming  the observed X-ray flux of the central source to be $\sim 3\%$ of the intrinsic X-ray flux, we could estimate the intrinsic X-ray flux to be $\approx{10^{41}} {\rm~ergs~ {s}^{-1}}$.  Assuming the X-ray luminosity to constitute $\sim 10\%$ of the bolometric luminosity ($L_{bol}$) and an Eddington ratio of $L_{bol}/L_{Edd} \sim  0.01$, we could estimate an Eddington luminosity of $L_{Edd}\sim{10^{44}}{\rm~ergs~ {s}^{-1}}$ and thus a central black hole mass of $\approx{10^6} M_\odot$. Since the X-ray flux (0.2-12 keV) of the source `B' in the ring like structure of the galaxy is also of the same order as the central source of the galaxy, it would also result in a black hole mass of $\approx{10^6} M_\odot$ if it is a type 2 Compton-thick AGN. If the source B is either Compton-thin or unabsorbed, the black hole mass is expected to be much lower. This suggests that there might a pair of accreting SMBHs,  the pair might have been formed due to possible merger of two galaxies which also likely led the formation of the ring-like structure and the additional extension of the diffused emission in the upper-right region in Figure~\ref{fig:int_srcs_out_1} (right).
    
    In the NUV image of the galaxy we clearly see extended diffused emission in the outer regions with additional extended emission in the upper-right region. This additional extended emission also appears to have a blob-like structure in the upper regions. The presence of two X-ray sources, if they are accreting SMBHs, and the extended diffuse emission in the upper-right region may have resulted from the collision of two  galaxies in which a dwarf galaxy possible passed through the central disk of the bigger spiral galaxy almost face-on and removed some of the gas in the central region; some fraction of the dwarf galaxy mass possibly came out in the upper-right direction, thus forming the observed additional extended region with a blob. Such collisions are thought to result in ring-like structures \citep{1976ApJ...209..382L}. This interaction may have also caused enhanced star formation in the annular rim due to shocks caused by the passing through of the small galaxy.  If this picture is correct, then we expect the extended UV emitting region to be outflowing which can be verified with long-slit and/or integral field spectroscopy of this galaxy.

    \begin{figure*}
    \epsscale{0.6}
    \plotone{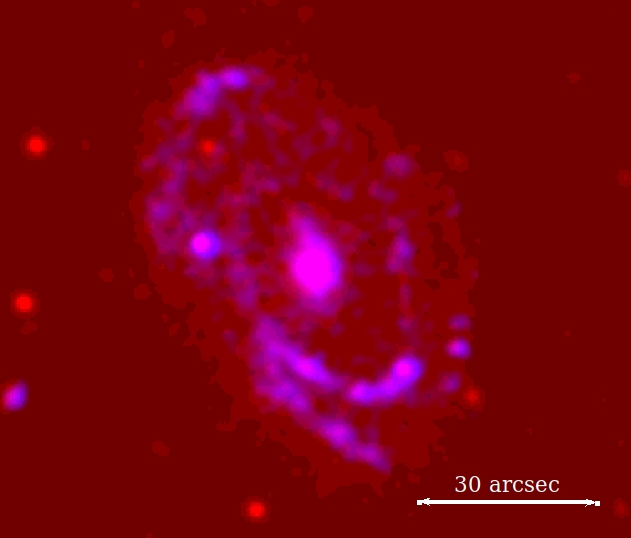}
    \caption{Two colour image of NGC 5302 constructed by showing NUV band image in red colour and FUV band image in blue colour. The galaxy is seen with a UV bright central core along with  prominent spiral arms that are bright in both FUV and NUV. A clear bifurcation can also be seen in one of the spiral arms (the arm in the lower left region of the image)
    \label{fig:int_srcs_out_2}}
    \end{figure*} 
    \item NGC~5302: The galaxy shown in the two-color FUV/NUV image of Figure~\ref{fig:int_srcs_out_2} is located at sky position ($\alpha (J2000) = 207.20681^{\circ}$, $\delta (J2000) =  -30.51099 ^{\circ}$). This galaxy is identified as NGC~5302 and is classified as a lenticular galaxy. The UV emission from the galaxy clearly shows a ring-like structure or tightly bound spiral arms. The inner ring-like structure or the spiral arm in the lower-left region appears to bifurcate (see Figure~\ref{fig:int_srcs_out_2}). Such bifurcating structures are extremely rare, and it is not clear how such structures can arise.   The galaxy is also seen to possess a strong FUV  and NUV emitting central region. NGC~5302 is known to be an emission line galaxy with a spectroscopic redshift of 0.01185. We measured  the UV flux of the central source to be $19.6 \pm 0.4 \times 10^{-17}{\rm~ergs~cm^{-2}~s^{-1}}$\AA$^{-1}$ in NUV band and $17.2\pm0.9 \times 10^{-17}{\rm~ergs~cm^{-2}~s^{-1}}$ \AA$^{-1}$ in FUV band. We found an X-ray source in the \xmm{} Slew Survey catalogue that is located within $\sim 1$ arcsec of the central UV source. The maximum X-ray flux is observed to be $1.3\pm0.4 \times 10^{-13}{\rm~ergs~cm^{-2}~s^{-1}}$and the luminosity of the central X-ray source is found to be $ 3.6 \pm  1.3 \times 10^{40} {\rm~ergs~s^{-1}}$. The presence of emission lines and X-ray emission from the nuclear regions suggest this galaxy to host a type 2 AGN. To find an approximate value of the central black hole mass of NGC 5302 , we repeated the same type of calculations we did for NGC 5298. Considering the observed X-ray (0.2-12 keV) flux to be $\approx{3\%}$ of the intrinsic X-ray flux (0.2-12 keV) which is assumed to be $\approx 10\%$ of bolometric flux and assuming an Eddington ratio of 0.01 we estimated the central black hole mass to be $\approx{10^{7}} M_\odot$. 
\end{enumerate}

\section{Summary \& Conclusions}\label{sect:summary}
In this work, we analyzed FUV and NUV data on IC~4329A field observed with \astrosat{}/UVIT in the broadband filters F154W and N245M. We performed astrometry and photometry, and generated a catalogue. We characterised the sources and investigated their UV variability. The main findings of our work are as follows.
\begin{enumerate}
    \item This study has resulted in an UVIT source catalogue of the field around IC~4329A field based on deep AstroSat/UVIT observations.
    \item We found 4437 sources above the SNR of 5 in our NUV field, out of which 456 sources have FUV counterparts, 651 sources have optical (\gaia{}) counterparts and 97 sources have X-ray (\xmm{}) counterparts. Only 10 sources have counterparts in all four bands.
    \item Out of these 4437 sources 1871 are point-like and the rest 2566 sources are extended.
    \item We reached a 5$\sigma$ detection limit of 26.0 and 25.1 mag in the NUV and the FUV band, respectively, with an exposure time of 82.9 ks in the N245M and 92.2 ks in the F154W filter. 
    \item Investigating color-color diagram using three bands: the NUV, the FUV and \gaia{} G band we found that 8 sources in our field that have counterparts in all of these three bands but have no significant proper motion are likely to be type~1 AGNs. We have also identified two of these 8 sources in Million Quasars catalogue.
    \item  By measuring $F_{\rm{var}}$ of the point sources in our catalogue we found 28 sources that were variable above $2.5 \sigma $ significance level in NUV band. Out of these 28 sources 3 were also found to be variable above $2.5 \sigma $ significance level in FUV band.
    \item We have found a number of extended sources with peculiar morphology. These include (i) a pair of bright FUV and NUV emitting nuclei in the galaxy LEDA~719364, most likely due to two accreting SMBHs though compact star forming regions can also create such structures, (ii) a ring-like structure in the galaxy LEDA~720905, (iii) complex structure in the galaxy FLASH~J134835.63-302441.5 possibly due to merging of two small galaxies, (iv) multiple blobs in the central regions of the spiral galaxy LEDA~89764, (v) a faint galaxy identified with WISEA~J134954.43-301349.0 that shows complex UV structure possibly due to merger/interaction, (vi) presence of ring-like structure, additional extended emission on one side, and two point-like central sources in NGC~5298, (vii) Unusual ring-like structure or bifurcating spiral arms and a strong nuclear UV source in NGC~5302.

    \item With our NUV band data, we established that the two white dwarf candidates in our field we could find by cross-matching white dwarf catalogue based on GAIA EDR3 data are not really white dwarfs.
\end{enumerate}

     We lacked optical spectroscopic or multi-band optical photometric data on our sources. We could not construct spectral energy distributions,  which could have given us further information about the true nature of these sources.

     The field we have studied harbours many new UV bright sources that had not been identified previously. We could find the optical and X-ray counterparts for a small number of sources in our catalogue. Thus, a large number of sources in our catalogue are being reported for the first time. Future deep optical spectroscopy and multi-wavelength follow-up observations of the field will reveal a wealth of information regarding the nature of the sources including variability properties.
     As our catalogue lists bright and interesting sources of a part of the UV sky which has not been explored previously in detail, it can serve as a source of information for future UV missions that are lined up for launching, for example ULTRASAT (The Ultraviolet Transient Astronomy Satellite) \citep{2024ApJ...964...74S}, UVEX (The Ultraviolet Explorer) \citep{2021arXiv211115608K} and QUVIK (The Quick Ultra-Violet Kilonova surveyor) \citep{2024SSRv..220...11W}. ULTRASAT, planned to launch in 2026, offers a better sensitivity (22.5 mag, 5$\sigma$, ar 900s) and much larger field of view ($204^\circ$) in NUV band (2300-2900 \AA) than UVIT and aims to do time domain survey of transient and variable NUV sources. UVEX, which is planned to launch in 2028, will have two detectors operating simultaneously in FUV (1300-1900 \AA) and NUV (2030-2700 \AA) band to perform photometric, spectroscopic and timing studies of the UV sky. QUVIK is a proposed UV photometry mission which will also offer simultaneous imaging in NUV (2600-3600 \AA) and FUV (1400- 1900 \AA) band with an angular resolution of $\lesssim 2.5$ arcsec (in NUV), NUV sensitivity of 22 AB magnitude (at SNR of 5 for 1000s) and FUV sensitivity of 20 AB magnitude (at SNR of 5 in 1000s). These UV survey missions will further probe the variability and spectral classes of the sources including interacting/merging galaxies in our catalogue. As the spatial resolution of UVIT is better than these future detectors, our catalogue can not only serve as a guidance map to find out interesting bright sources for further studies but also serve as an additional data source for these future UV missions.

\begin{acknowledgments}
This publication utilizes the data from Indian Space Science Data Centre (ISSDC) of the\textit{AstroSat} mission of the Indian Space Research Organisation (ISRO). This publication uses UVIT data processed by the payload operations center at Indian Institute of Astrophysics (IIA). The UVIT is built in collaboration between IIA, Inter-University Centre for Astronomy and Astrophysics (IUCAA), Tata Institute of Fundamental Research(TIFR), ISRO, and the Canadian Space Agency (CSA).  UVIT data were reprocessed by CCDLAB pipeline. The raw data we used for our analysis can be downloaded from \astrosat{} data archive: \url{https://astrobrowse.issdc.gov.in/astro_archive/archive/Home.jsp}. This research has made use of the SIMBAD database, operated at CDS, Strasbourg, France. This research has made use of the NASA/IPAC Extragalactic Database (NED), which is funded by the National Aeronautics and Space Administration and operated by the California Institute of Technology. This research has made use of data and/or software provided by the High Energy Astrophysics Science Archive Research Center (HEASARC), which is a service of the Astrophysics Science Division at NASA/GSFC.
\end{acknowledgments}

\vspace{4mm}
\facilities{\astrosat (UVIT), \gaia, \xmm, NRAO VLA}


\software{CCDLAB \citep{2017PASP..129k5002P, 2021JApA...42...30P},  
          Source Extractor \citep{1996A&AS..117..393B}
          }

\bibliography{IC4329a_paper}
\bibliographystyle{aasjournal}



\end{document}